\documentclass[10pt, twocolumn]{iopart}

\usepackage{comment}
\usepackage{xcolor}
\usepackage{array}
\usepackage{arydshln}
\expandafter\let\csname equation*\endcsname=\relax 
\expandafter\let\csname endequation*\endcsname=\relax 
\usepackage{amsmath}
\usepackage{amssymb}
\usepackage{subcaption}
\usepackage{gensymb}
\usepackage{cancel}
\usepackage[margin=0.6in]{geometry}
\usepackage{makecell}
\usepackage{graphicx,booktabs}
\usepackage{multicol}
\usepackage{url}

\usepackage[separate-uncertainty=true]{siunitx}
\DeclareSIUnit\torr{torr}

\begin{document}

\twocolumn[
  \begin{@twocolumnfalse}
    \title{Empirical impact of near-separatrix plasma and neutral transport on the pedestal in the transition between EDA and ELMy H-modes on Alcator C-Mod}

\author{M.A. Miller$^{1,2}$,
        J.W. Hughes$^1$,
        S. Saarelma$^3$,
        T. Eich$^4$,
        J. Dunsmore$^1$,
        J. Han$^1$,
        P. Manz$^{5,6}$,
        J.W. Connor$^3$,
        G.R. Tynan$^{1,7}$,
        A.E. Hubbard$^1$,
        A. Ho$^1$,
        T. Body$^4$,
        D. Silvagni$^2$,
        O. Grover$^2$,
        S. Mordijck$^8$,
        E.M. Edlund$^9$,
        B. LaBombard$^1$,
        M. Wigram$^1$,
        A. Cavallaro$^1$
        }

\address{$^1$MIT Plasma Science and Fusion Center, Cambridge, MA 02139, USA
}
\address{$^2$Max-Planck-Institut für Plasmaphysik, Boltzmannstraße 2, D-85748, Garching, Germany
}
\address{$^3$United Kingdom Atomic Energy Authority, Culham Campus, Abingdon OX14 3DB, United Kingdom of Great Britain and Northern Ireland}
\address{$^4$Commonwealth Fusion Systems, Devens, MA 01434, USA
}
\address{$^5$Institute of Physics, University of Greifswald, Felix-Hausdorff-Str.6, Greifswald, 17489, Germany
}
\address{$^6$Max-Planck-Institut für Plasmaphysik, Wendelsteinstr. 1, 17491 Greifswald, Germany
}
\address{$^7$University of California, San Diego, CA 92093, USA
}
\address{$^8$William \& Mary, Williamsburg, VA 23188, USA
}
\address{$^9$Department of Physics, SUNY Cortland, Cortland, New York 13045, USA
}
\ead{millerma@mit.edu}

\maketitle
    \label{sec:abstract}

\begin{abstract}
The transition between the ELMy H-mode and the EDA H-mode is studied on Alcator C-Mod using an experimental database and recently-developed predictive pedestal models. High-resolution Thomson scattering measurements are used to inspect both the separatrix and pedestal operational space using two independent fit functions, each tailored to analyze either the pedestal or the separatrix and near-SOL profiles. The pedestal density, $n_{e}^\mathrm{ped}$, and the separatrix density, $n_{e}^\mathrm{sep}$ are compared to main chamber neutral measurements. $n_{e}^\mathrm{ped}$ is sensitive to neutral sources only in the ELMy H-mode regime and not in the EDA H-mode regime. Density fluctuation spectra reveal that quasi-coherent structures become stronger at higher densities and more coherent in the EDA relative to the inter-ELM phases of ELMy H-modes, before weakening again at the highest values of $n_{e}^\mathrm{ped}$. The Saarelma-Connor pedestal density prediction model is validated for ELMy H-modes up to $n_{e}^\mathrm{ped} = 2.0 \times 10^{20}$ m$^{-3}$. An additional transport channel driven by resistive ballooning modes (RBM), $D_\mathrm{RBM}$, scaling directly with $\alpha_{t}$, a collisional turbulence control parameter, and inversely with $k_\mathrm{RBM}^{2}\hat{q}_\mathrm{cyl}$, the product of the square of the characteristic RBM turbulence wavenumber and the cylindrical safety factor, is shown to improve the prediction for EDA H-modes, finding good model agreement up to $n_{e}^\mathrm{ped} = 3.0 \times 10^{20}$ m$^{-3}$. For comparison, EPED scans in $n_{e}^\mathrm{ped}$ are then performed at three values of $n_{e}^\mathrm{sep}/n_{e}^\mathrm{ped}$. Increasing this ratio moves the peeling-ballooning branch transition to lower $n_{e}^\mathrm{ped}$, increasing $p^\mathrm{ped}$ in the peeling branch and decreasing it in the ballooning branch. Agreement is found for large ELM H-modes. SPARC pedestal density predictions for an ELMy and an EDA/QCE-like H-mode are performed using the extension to the standalone Saarelma-Connor model and are found consistent with earlier assumptions used in EPED modeling. Inclusion of $D_\mathrm{RBM}$ significantly weakens the density gradient near the separatrix, lowering $n_{e}^\mathrm{ped}$ by approximately 20\%.
\end{abstract}
  \end{@twocolumnfalse}
]

\section{Introduction}
\label{sec:intro}

In fusion energy research, finding a solution that integrates a hot core plasma with a cold edge plasma is of high priority. Doing so is predicated on solid understanding and predictive capability of the region of the plasma that connects these two very different plasma states. In the high-confinement mode (H-mode) \cite{asdex_team_h-mode_1989}, this region is unsurprisingly characterized by steep gradients. It is called the pedestal, and understanding the mechanisms that determine its structure is the subject of a large body of research \cite{fenstermacher_progress_2025}. Understanding the pedestal requires a multi-disciplinary approach -- it is the intersection of the plasma physics of microscale turbulent instabilities and macroscopic magnetohydrodynamic (MHD) modes with that of atomic physics determining ionization, charge exchange, and recombination processes. When gradients in the pedestal become too steep, the edge plasma may reach a soft, non-disruptive limit in the form of an edge localized mode (ELM) \cite{leonard_edge-localized-modes_2014}, expelling a large fraction of the plasma stored energy onto open field lines and directly onto the plasma facing components. As the ELM energy fluence to the divertor is expected to scale with both the electron pressure and the device major radius \cite{eich_elm_2017}, both of which will increase in next-step devices and eventually reactors, any solution featuring the most pernicious type of ELM, a Type-I ELM, will not be tolerated \cite{leonard_impact_1999, kuang_divertor_2020}. A number of alternative scenarios to the Type-I ELMy H-mode exist, and understanding access to these, avoidance of ELMs, and optimization of their performance is of great interest \cite{viezzer_prospects_2023}.

This paper carries out detailed experimental analysis and modeling of one such confinement scheme, the enhanced D$_{\alpha}$ (EDA) H-mode on Alcator C-Mod, building on recently published analysis of the transition between the EDA and the ELMy H-mode at the separatrix \cite{miller_determination_2025}. The EDA H-mode was discovered and routinely accessed on Alcator C-Mod \cite{greenwald_characterization_1999, greenwald_cmod_2014}. It has since been found and is actively studied on AUG \cite{gil_stationary_2020, gil_eda_2025} and DIII-D \cite{macwan_elm-free_2024}, and most recently on JET \cite{gil_high-confinement_2025}. It exists at high edge density, which is attractive from the point of view of divertor survivability. Crucially, it is unknown whether it can exist at the high temperatures expected in the pedestal on future devices. The EDA H-mode is thought to be closely related to the quasi-continuous exhaust (QCE) regime \cite{faitsch_broadening_2021}, a small-ELM regime also favored at high density and high shaping \cite{dunne_quasi-continuous_2024}. While the QCE has not been identified in the the current EDA-only dataset, the analysis in this paper makes strides in understanding avoidance of Type-I ELMs at high density, helping clarify the reactor-relevance of such regimes, be they EDA or QCE. This is done using findings about the relevant physical mechanisms that determine the edge profile at high density and using them to make additions to and suggestions for future improvements in pedestal modeling. The paper then applies the findings from this analysis to make predictions for the edge pedestal of the SPARC tokamak \cite{creely_overview_2020}. 


Expanding on recent separatrix analysis of the ELMy-EDA H-mode transition \cite{miller_determination_2025}, Section \ref{sec:exp_profs_neutrals} begins by presenting data taken with the same Thomson scattering diagnostic, now considering how plasma quantities in this dataset evolve throughout the edge more generally, with a focus on the pedestal and its gradients. Earlier analysis provided evidence that there is key physics of interest at the separatrix for this transition, with implications for types of turbulence active in the edge and possibly for the suppression of large ELMs \cite{faitsch_analysis_2023, miller_determination_2025, li_exploring_2025}. The current work shows that the same driving parameters may also influence the type of H-mode achieved when estimated at the pedestal top. By considering an additional measurement of neutral pressure in the main chamber, it finds that in the EDA H-mode, the density profile is strongly influenced by particle transport in the pedestal, whereas in the ELMy H-mode, it is much more easily modified by changes to fueling sources. In in Section \ref{sec:fluctuations}, analysis of fluctuation power spectra from the phase contrast imaging (PCI) diagnostic is carried out. The evolution of the quasi-coherent mode (QCM) characteristic of EDA H-modes is tracked and compared with the inter-ELM fluctuation spectra of ELMy H-modes, and its strength is compared to physical parameters extracted from the plasma profiles. This analysis provides an important experimental comparison from which to build understanding on the relevant fluctuation-driving micro-instabilities in the pedestal of these different H-modes.

The second half of this paper focuses on testing and expanding on state-of-the art models for pedestal prediction. In particular, Section \ref{sec:density_model_validation} examines the recently-developed Saarelma-Connor pedestal density prediction model \cite{saarelma_testing_2023} and Section \ref{sec:eped} focuses on the well-validated EPED model \cite{snyder_development_2009, snyder_eped_2012}. The former was developed using JET data and has now also been tested on MAST-U and AUG \cite{saarelma_density_2024}. It includes both neutrals and plasma transport models, offering a testbed on which to scrutinize the interplay between these two populations in the edge, as well as the role each plays in determining the pedestal density in different H-mode regimes. EPED scans are then performed for a large range of pedestal densities and at different values of the ratio of the separatrix to pedestal density. Experimental pedestal data are then revisited to provide a qualitative comparison for the evolution of the pedestal across different H-mode types with expectations from the EPED model. Analysis of the EDA H-mode and its fueling and plasma transport characteristics helps to provide an understanding of how one might develop an edge profile without large ELMs on a next-generation device like SPARC. Section \ref{sec:sparc_prediction} concludes the body of this paper by leveraging these findings about pedestal transport, as well as a boundary condition consistent with predictions from the separatrix operational space (SepOS) model \cite{eich_separatrix_2021}, to make predictions for the edge density profile on SPARC for both the primary reference discharge (PRD) scenario \cite{Rodriguez-Fernandez_2022, body_sparc_nodate} and for a recently-proposed high-density scenario. Predictions for the latter suggest that at the proposed density, particle transport just inside the separatrix may act to strongly limit the pedestal density gradient, with the benefit of possibly avoiding Type-I ELMs. Section \ref{sec:conclusions} closes with concluding remarks and proposes work to improve physical understanding and predictive capabilities of regimes free of Type-I ELMs.

\section{Experimental profiles and the influence of neutrals}
\label{sec:exp_profs_neutrals}

\begin{figure*}
\centering
\includegraphics[width=1.7\columnwidth]{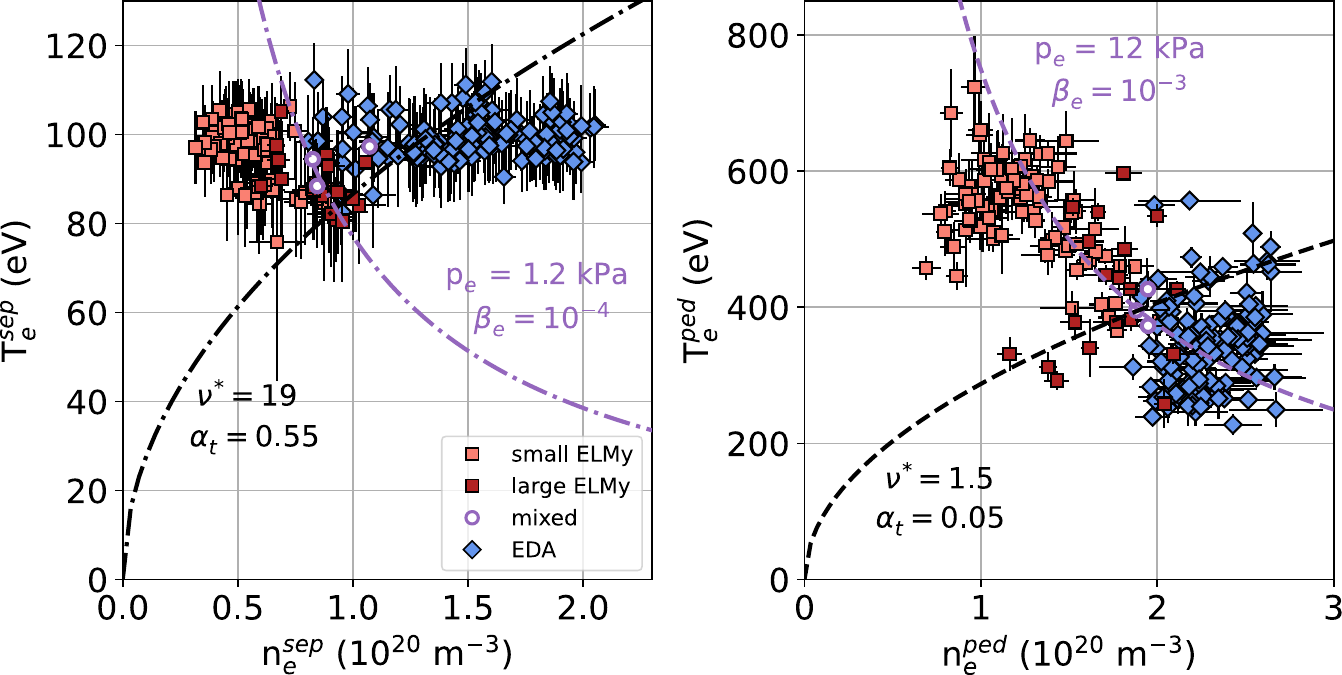}
\caption{Operational space in terms of $T_{e}$ and $n_{e}$ at the separatrix (left) and pedestal top (right), showing discharges with small ELMs (light red squares), large ELMs (dark red squares), mixed EDA and ELMy (purple circles), and only EDA (blue diamonds). At left, dash-dotted black and purple lines show $\alpha_{t}$ and $\beta_{e}$ contours at the separatrix from \cite{miller_determination_2025}. At right, the dashed black line shows the contour of constant collisionality, $\nu^{*} = 1.5$ ($\alpha_{t} = 0.05$), and the dashed purple line shows the contour of constant electron pressure, $p_{e} = 12$ kPa ($\beta_{e} = 10^{-3}$).}
\label{fig:sepos_vs_pedos}
\end{figure*}

All data presented here come from one run day featuring experiments to study fluctuations in the ELMy H-mode \cite{diallo_observation_2014, diallo_correlations_2015}. This run day features plasmas only in a non-standard shape, with low upper triangularity relative to the more typical C-Mod shape, and with the outer strike point in the lower divertor slot instead of on the vertical target. A comparison of this shape with the standard C-Mod shape is depicted in Figure 1 of \cite{hughes_pedestal_2013}. Throughout the run day, the programmed density in the L-mode was reduced. Particle control, as with many H-modes on C-Mod, was achieved via gas puff feedback control prior to the H-mode transition, after which no additional fueling was needed to sustain the H-mode. At the beginning of the run day, plasmas were in the EDA regime, the typical operating regime for H-modes on C-Mod. Below a particular value, however, plasmas transitioned away from the EDA and ELMs appeared. On Alcator C-Mod, ELMs were only observed at a particular plasma shape and at low collisionality \cite{hughes_observations_2002, hughes_pedestal_2013}. Discharges were all at fixed plasma current, $I_{P} = 0.9$ MA, toroidal magnetic field, $B_{t} = 5.6$ T, elongation, $\kappa = 1.6$, and average triangularity, $\delta = 0.5$. They were all in lower single null (LSN), with the upper divertor cryopump warm and the drift direction pointing in the favorable direction, towards the lower X-point. The H-mode transition was triggered via injection of ion cyclotron range of frequencies (ICRF) power ranging between 2 -- 3 MW, with most EDA H-modes requiring close to 3 MW and most ELMy H-modes only slightly above 2 MW. Classification of data points in the current paper follows that described in Section 6 of \cite{miller_determination_2025}, and the full range of parameters spanned in the dataset can be found in the last column of Table 1 in the same.  


The primary diagnostic used to analyze the pedestal of these H-modes is the edge Thomson scattering (ETS) diagnostic on C-Mod \cite{hughes_high-resolution_2001, hughes_thomson_2003}, which measured electron density, $n_{e}$, and electron temperature $T_{e}$ throughout the edge. Local values of these profiles and their gradients are estimated using two different fitting approaches. The first approach makes use of a modified hyperbolic tangent (mtanh), as detailed in \cite{miller_enhanced_2025}. This fit function captures the characteristic shape of H-mode pedestals on Alcator C-Mod, returning useful quantities related to the structure of the pedestal, like its height and width. Previous work has shown that on C-Mod, ELM filtering does not strongly affect reported pedestal values except for large ELMs \cite{walk_characterization_2012}. Subsequent analysis showed that TS $p_{e}$ measurements at the pedestal top are generally stationary after the first 30\% of the ELM cycle \cite{hughes_pedestal_2013}. As such, ELM filtering is done for 30\% of the cycle, leaving only TS data taken in the last 70\%. The second fitting approach is designed to capture the decay of kinetic profiles across the separatrix in order to locate the separatrix with higher fidelity and in doing so, estimate characteristic gradient scale lengths at this radial location and into the near scrape-off layer (SOL). Separatrix position identification is done using the two-point model with the $T_{e}$ gradient scale length, $\lambda_{T_{e}}$, used to estimate the heat flux width, $\lambda_{q}$ under the assumption of Spitzer-Härm transport. Details of this approach using a simple exponential decay fitting function are reported in \cite{miller_determination_2025}. No ELM filtering is done for separatrix fits as the ETS data near the separatrix are largely unaffected by ELMs on C-Mod \cite{hughes_pedestal_2013}.

\subsection{The edge plasma operational space}
\label{subsec:edge_os}

Figure \ref{fig:sepos_vs_pedos} shows the operational space (OS) in $n_{e}$ and $T_{e}$ for this entire dataset at two different locations in the edge profile. At left, it shows separatrix quantities, estimated using the exponential decay iterative fitting approach. The separatrix OS for these discharges has been shown in Figure 8 of \cite{miller_determination_2025}, although the dataset is slightly reduced compared to that work, including only time windows that also have sufficiently good ETS measurements up to the pedestal top. At right, Figure \ref{fig:sepos_vs_pedos} shows the OS at the pedestal top, $n_{e}^\mathrm{ped}$ and $T_{e}^\mathrm{ped}$, defined using the mtanh fit. Note that the two quantities are not in general radially co-located, as the two profiles are not guaranteed to have either the same center position or the same width. An outward shift of $n_{e}$ relative to $T_{e}$ typically exists between the profiles, in particular for EDA discharges \cite{hughes_edge_2007, miller_enhanced_2025}.

The figures each show two other curves -- contours of constant $\alpha_{t}$ and constant $\beta_{e}$. $\alpha_{t}$ is a collisionality-like turbulence control parameter that characterizes the competition between drift-wave and interchange turbulence \cite{eich_turbulence_2020}. For typical C-Mod densities, $O(10^{20})$, $\alpha_{t} = 2.98 \times 10^{-18} R_\mathrm{geo}\hat{q}_\mathrm{cyl}^{2} \frac{n_{e}}{T_{e}^{2}} Z_\mathrm{eff}$, where, $R_\mathrm{geo}$ is the geometric major radius, taken to be equal to the device major radius, $R_{0}$, and $\hat{q}_\mathrm{cyl}$ is the cylindrical safety factor, calculated according to $\hat{q}_\mathrm{cyl} = \frac{B_{t}}{B_{p}} \times \frac{\hat{\kappa}}{R_\mathrm{geo}/a_\mathrm{geo}}$, with $a_\mathrm{geo}$ the geometric minor radius, taken equal to the device minor radius, and $\hat{\kappa}$ is the effective elongation, defined in \cite{eich_separatrix_2021} in terms of triangularity, $\delta$, and geometric elongation, $\kappa_\mathrm{geo}$. The second curve, $\beta_{e}$, is the electron beta, given by $\beta_{e} = \frac{p_{e}}{B^{2}/2\mu_{0}}$, where $p_{e}$ is the electron pressure, and $B$ is the total magnetic field. For the fixed $B_{t}$ and $I_{P}$ in this dataset, constant $\alpha_{t}$ and $\beta_{e}$ curves amount to collisionality isolines and pressure isobars, respectively.

The $\alpha_{t}$ and $\beta_{e}$ isolines shown for the separatrix OS correspond to values proposed in \cite{miller_determination_2025} as possible criteria for the transition between the two types of H-modes. Dashed curves on the right, show two other values of $\alpha_{t}$ and $\beta_{e}$ that organize the pedestal OS. The hotter (and somewhat denser) pedestal region reaches considerably larger $p_{e}$ (or $\beta_{e}$) values and considerably lower $\nu^{*}$ (or $\alpha_{t}$) values. Unlike at the separatrix, a boundary in collisionality, here $\nu^{*} = 1.5$, appears more appropriate than a boundary in $p_{e}$. Instead, many discharges reach similar $p_{e}^\mathrm{ped} \approx 12$ kPa. Many ELMy, as well as some EDA, discharges have $p_{e} < 12$ kPa, whereas a number of especially hot EDA H-modes reach $p_{e} > 12$ kPa. Although many EDA H-modes and ELMy H-modes have similar $p_{e}^\mathrm{ped}$, the former reach these pressures largely due to high $n_{e}^\mathrm{ped}$ and the latter do so due to high $T_{e}^\mathrm{ped}$. These results are consistent with previous analysis comparing the two regimes on Alcator C-Mod \cite{hughes_pedestal_2013}. Discharges lying along the isobar at 12 kPa can be either EDA or ELMy. Note that most discharges with large ELMs lie along this isobar (despite scatter), although many discharges with small ELMs discharges do as well. Constant pedestal pressure is characteristic of ballooning-limited ELMy profiles. Limits imposed by coupled-peeling ballooning modes on $p_{e}^\mathrm{ped}$ will be discussed in greater detail in Section \ref{sec:density_model_validation}.


\begin{figure}
\centering
\includegraphics[width=\columnwidth]{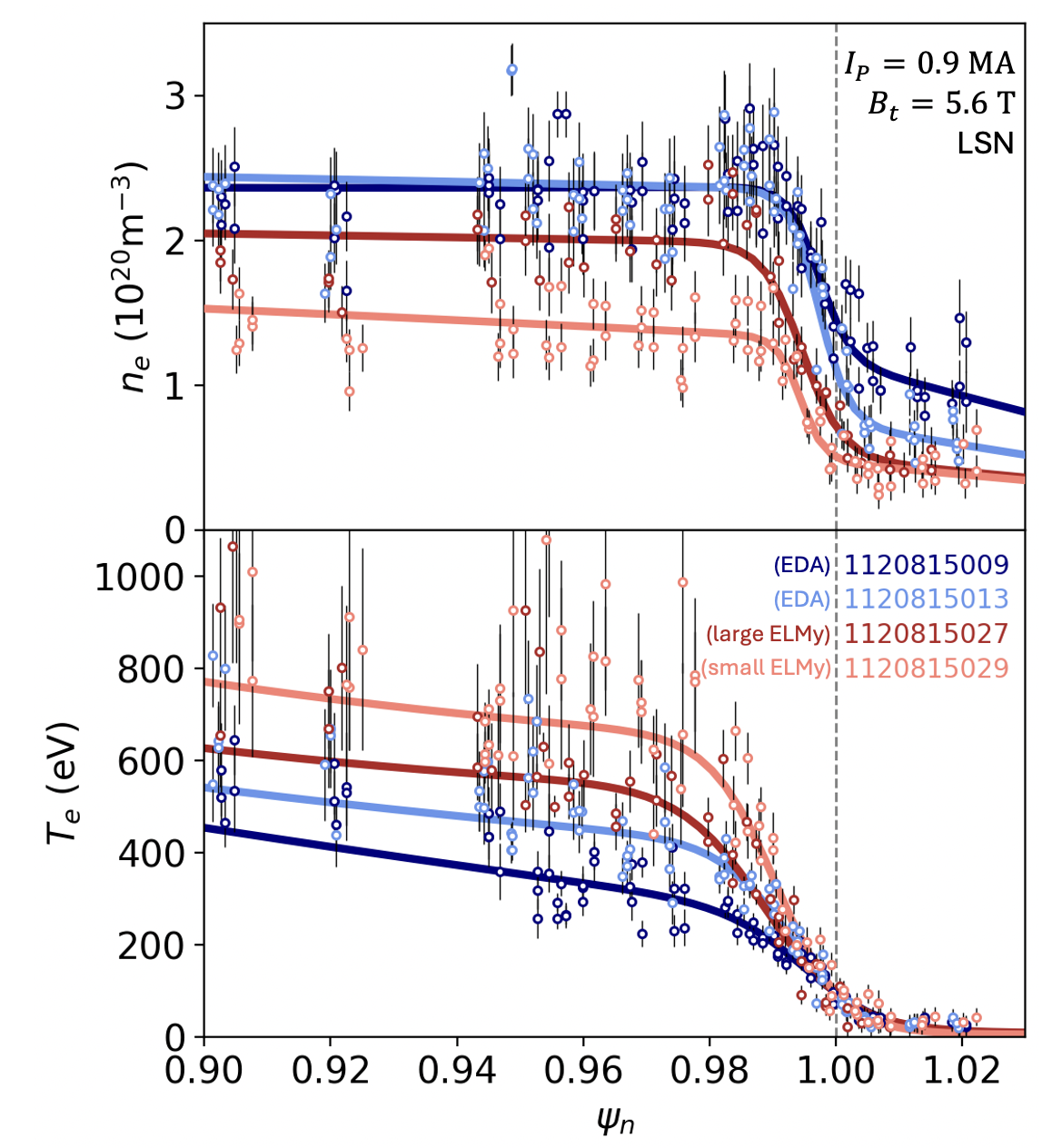}
\caption{Profiles of $n_{e}$ (top) and $T_{e}$ (bottom) against $\psi_{n}$ showing a high $n_{e}$ EDA H-mode in dark blue, a medium $n_{e}$ EDA H-mode in light blue, a medium $n_{e}$ large ELMy discharge as dark red, and a low $n_{e}$ small ELMy discharge in light red.}
\label{fig:elmy_eda_profiles}
\end{figure}

\begin{figure}
\centering
\includegraphics[width=0.9\columnwidth]{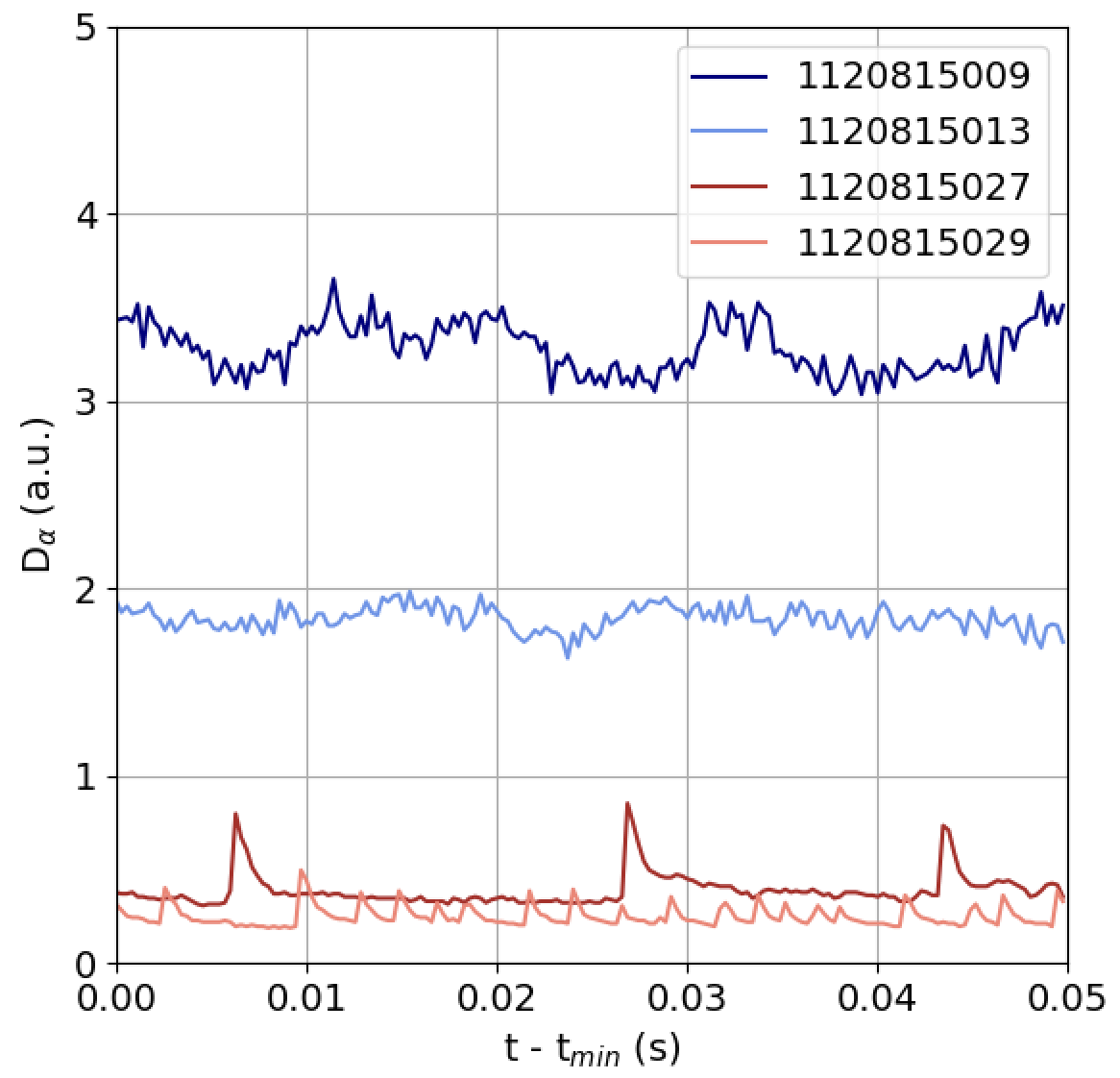}
\caption{D$_{\alpha}$ emission as a function of time from beginning of time window, $t_\mathrm{min}$, for selected discharges from Fig \ref{fig:elmy_eda_profiles}.}
\label{fig:dalpha_traces}
\end{figure}

Figures \ref{fig:elmy_eda_profiles} and \ref{fig:dalpha_traces} provide a closer look at experimental data from four characteristic discharges spanning this OS, all in the same non-standard LSN shape. The former shows typical $n_{e}$ and $T_{e}$ profiles and the latter shows typical time traces of D$_{\alpha}$ emission in each regime -- two ELMy H-modes in shades of red and two EDA H-modes in shades of blue. EDAs have elevated $n_{e}$ and lowered $T_{e}$ relative to the ELMy profiles, as well as greater D$_{\alpha}$ emission. The EDA in dark blue, produced earliest in the run day at the highest particle content also has the highest separatrix density, $n_{e}^\mathrm{sep}$, although not necessarily a higher $n_{e}^\mathrm{ped}$ than the EDA shown in light blue. Indeed, as the programmed L-mode density is reduced, it is initially $n_{e}^\mathrm{sep}$, and more generally the scrape-off layer (SOL) density, that is reduced, as well as the mean level of collected D$_{\alpha}$ light. Further decreases in the programmed L-mode density then spur the transition to the ELMy H-mode, yielding slightly lower $n_{e}$ throughout the edge and somewhat higher $T_{e}$. Here, the D$_{\alpha}$ emission is considerably lower than in both EDA H-modes, and large spikes in the emission characteristic of large ELMs are visible. Finally, as the programmed L-mode density is reduced further, the profiles in light red are achieved. $n_{e}^\mathrm{ped}$ is reduced significantly and $T_{e}^\mathrm{ped}$ grows to over 600 eV. The average level of D$_{\alpha}$ light drops somewhat, but more notably, the size of the ELM spikes decreases while their frequency increases, relative to the time traces in dark red. Note that while large ELM discharges on C-Mod have received considerable attention \cite{TERRY2007994, walk_characterization_2012, hughes_pedestal_2013, diallo_correlations_2015, snyder_high_2019}, these small ELM discharges on C-Mod have remained largely unanalyzed. They exist at low collisionality and may be more akin to a grassy-ELM regime \cite{wang_grassy_2021} than the QCE, for example, which is instead favored by high edge collisionality and features a broad SOL density shoulder \cite{faitsch_broadening_2021}.

Though not explicitly shown, $p_{e}^\mathrm{ped}$ is lowest for the high-density EDA H-mode in dark blue and the low-density ELMy H-mode in light red. They both suffer from reduced $n_{e}^\mathrm{ped}$ or $T_{e}^\mathrm{ped}$, respectively. This drops $p_{e}^\mathrm{ped}$ below the 12 kPa isobar shown in Figure \ref{fig:sepos_vs_pedos}, to which the two other profiles (relatively lower density EDA H-mode in light blue and relatively higher density ELMy H-mode in dark red) belong. $T_{e}^\mathrm{sep}$ does not vary much, in part by construction, as profiles are shifted to the separatrix to match the value of $T_{e}$ predicted by the two-point model, which only varies in this dataset (and weakly) with variation in the net power at the separatrix and $\lambda_{q}$. Despite the actual value of $T_{e}$ calculated from the two-point model, it is clear from the profiles in Figure \ref{fig:elmy_eda_profiles} that there is significantly more variation in $n_{e}$ than in $T_{e}$ in the vicinity of the separatrix, e.g. $0.99 < \psi_{n} < 1.01$. Furthermore, as the L-mode programmed density decreases, the absolute value of the density gradient, $|\nabla n_{e}|$, generally decreases and the density pedestal width, $\Delta_{n}$, slightly increases. The opposite is the case for the $T_{e}$ profile, with the absolute value of its gradient, $|\nabla T_{e}|$, increasing and its width, $\Delta_{T}$ decreasing as $n_{e}$ decreases. For very high density EDAs like that in dark blue, however, $|\nabla n_{e}|$ is lower than lower-density EDAs like that in light blue, even though they are at higher density, as a clear density shoulder emerges in the SOL, with $n_{e} > 10^{20}$ m$^{-3}$ in the near-SOL. Notably, $n_{e}$ at and inside the pedestal is nearly insensitive to changes in the SOL plasma -- $|\nabla n_{e}|$ throughout the edge adjusts to keep $n_{e}^\mathrm{ped}$ fixed. The implications for this phenomenology with view of the mechanisms that determine $n_{e}$, especially in the EDA, will be discussed in the following section. $\Delta_{T}$ appears slightly less variable for the ELMy discharges, but $\nabla T_{e}$ is observed to be systematically higher in the ELMy H-mode than in the EDA H-mode, yielding higher $T_{e}^\mathrm{ped}$. These relative changes in $\nabla T_{e}$ and $\nabla p_{e}$ may be important for identifying different instabilities that produce gradient-driven transport in pedestals, especially in ELMy H-modes.

\begin{figure*}
\centering
\includegraphics[width=1.7\columnwidth]{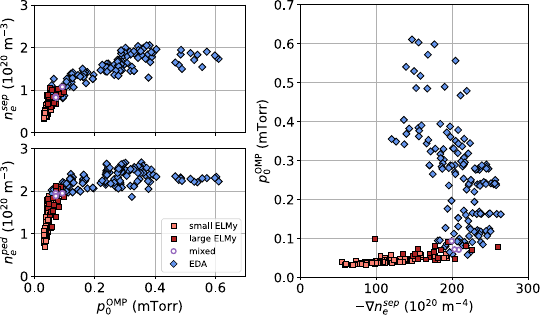}
\caption{$n_{e}^\mathrm{sep}$ (top, left) and $n_{e}^\mathrm{ped}$ (bottom, left) plotted against $p_{0}^\mathrm{OMP}$ and $p_{0}^\mathrm{OMP}$ plotted against $-\nabla n_{e}^\mathrm{sep}$ (right) for discharges with small ELMs (light red squares), large ELMs (dark red squares), mixed EDA and ELMy (purple circles), and only EDA (blue diamonds).}
\label{fig:ped_sep_neutrals_combined}
\end{figure*}

\subsection{The influence of fueling and transport on the density profile}
\label{sec:density_profile}

The remaining analysis in this section will focus on the $n_{e}$ profile, reporting on changes to its structure and how these are engendered by external actuators. It connects the observed stiffness in $n_{e}^\mathrm{ped}$ to changes to the transport properties of these H-modes. Figure \ref{fig:ped_sep_neutrals_combined} shows  information about the relationship between neutrals in the edge and achievable $n_{e}$ values. At left, the figure shows $n_{e}^\mathrm{sep}$ and $n_{e}^\mathrm{ped}$ at top and bottom, respectively, from the fits to the ETS profiles, plotted against $p_{0}^\mathrm{OMP}$, a measurement of neutral pressure made at the outer midplane (OMP) using a ratiomatic ionization gauge. At right, $p_{0}^\mathrm{OMP}$ is plotted against $\nabla n_{e}^\mathrm{sep} = n_{e}^\mathrm{sep}/\lambda_{n_{e}}^\mathrm{sep}$, evaluated from the exponential fit described earlier. A similar set of plots describing changes to fueling and particle transport was shown in Figures 4 and 5 of \cite{miller_enhanced_2025}. A key difference, however, is that the experiments presented in that work had the luxury of a Ly$_{\alpha}$ neutral emission measurement providing neutral densities and ionization rate profiles, while these do not. For C-Mod plasmas with a fixed geometry and relatively fixed SOL profiles, $p_{0}^\mathrm{OMP}$ serves as a  proxy for both the volumetric ionization rate, $S_\mathrm{ion}$, and the cross-field particle flux, $\Gamma_{\perp}$. While the mapping among these quantities empirically is unknown for the current dataset, this relationship is examined through simulations with a kinetic neutral code discussed in Section \ref{sec:density_model_validation} and detailed in \ref{sec:kn1d_sims}. Note that the pressure measured at the outer divertor with a baratron is anywhere between 60 -- 100 times higher than $p_{0}^\mathrm{OMP}$, and it trends approximately linearly with $p_{0}^\mathrm{OMP}$.


The striking similarities between these plots and those from \cite{miller_enhanced_2025}, as well as the quite abrupt transition in behavior of the plasma profiles at a critical value of $p_{0}^\mathrm{OMP}$, suggest important implications for the influence of fueling vs. transport in these different types of pedestals. The ELMy-EDA H-mode transition occurs at $p_{0} \approx 0.1$ mTorr, such that ELMy H-modes have $p_{0}^\mathrm{OMP} < 0.1$ mTorr, and EDA H-modes have $p_{0}^\mathrm{OMP} > 0.1$ mTorr. When the neutral source is low, $n_{e}^\mathrm{ped}$ appears sensitive to a change in that source. Indeed, $n_{e}^\mathrm{ped}$ can vary between $0.5 - 2.0 \times 10^{20}$ m$^{-3}$ in the ELMy H-mode. $n_{e}^\mathrm{sep}$ is nearly as sensitive to changes to $p_{0}^\mathrm{OMP}$, varying between $0.3 - 1.0 \times 10^{20}$ m$^{-3}$. As $p_{0}^\mathrm{OMP} \rightarrow 0.1$ mTorr, H-modes transition to EDA, which have a much broader range in $p_{0}^\mathrm{OMP}$ than ELMy H-modes, ranging from $0.1 - 0.6$ mTorr. Furthermore, the rate at which $n_{e}$ changes with $p_{0}$ varies considerably in this regime. $n_{e}^\mathrm{sep}$ grows more slowly with $p_{0}$, although it can continue to grow up to and even above $2.0 \times 10^{20}$ m$^{-3}$. $n_{e}^\mathrm{ped}$, on the other hand, is almost completely insensitive to changes to the source of neutrals. As $p_{0}^\mathrm{OMP}$ increases beyond 0.1 mTorr and $n_{e}^\mathrm{sep}$ continues to grow, $n_{e}^\mathrm{ped}$ does not, reaching a saturated value of $n_{e}^\mathrm{ped} \approx 2.5 \times 10^{20}$ m$^{-3}$ when $p_{0}^\mathrm{OMP} = 0.3$ mTorr, and even decreasing slightly as $p_{0}^\mathrm{OMP} \rightarrow 0.6$ mTorr. At the highest values of $p_{0}$, $T_{e}^\mathrm{ped}$ begins to decrease as well, leading to a drop in $p_{e}^\mathrm{ped}$ for these discharges below the 12 kPa isobar shown in Figure \ref{fig:sepos_vs_pedos}. These observations are consistent with earlier analysis of strictly EDA H-modes on C-Mod  \cite{miller_enhanced_2025}, as well as subsequent analysis of the same discharges suggesting that this phenomenology may be closely linked to rapid increases in both particle and heat transport associated with proximity to the H-mode density limit \cite{miller_thesis_2025}. The discharges in those studies were at the same $I_{P}$ and $B_{t}$, albeit in the more standard shape with the strike point in its more typical place on the outer vertical target and in a configuration closer to double null.

It should be noted that the dependence of $n_{e}^\mathrm{sep}$ on changes in neutral content is an active area of study in boundary physics. A recent multi-machine study including data from C-Mod, AUG, and JET, compared $n_{e}^\mathrm{sep}$ with divertor neutral pressure, $p_{0}^\mathrm{div}$, and proposed a semi-empirical scaling for the two parameters based in part from the reversed two-point model \cite{silvagni_crossmachine_2024}. Separately, ITER divertor gas throughput scans using SOLPS-ITER observed similar saturation of the upstream $n_{e}^\mathrm{sep}$ with increasing sub-divertor neutral pressure, a phenomenon it attributed to neutral compression and increasing SOL opaqueness \cite{park_assessment_2021}. Here, similar dependence of $n_{e}^\mathrm{sep}$ on $p_{0}$ is observed, noting of course the key difference that $p_{0}$ here is measured at the OMP rather than at the divertor, an observation which neglects possibly important and likely present asymmetries in the neutral fueling profile. The results shown here also differ in that they show the effect of changing neutral content on $n_{e}^\mathrm{ped}$, which very clearly has a different dependence, especially at the regime transition.


These changes can be visualized even more clearly when considering the right panel of Figure \ref{fig:ped_sep_neutrals_combined}, a makeshift analog of the $\Gamma-\nabla n$ plot shown in \cite{miller_enhanced_2025}. For ELMy discharges with $p_{0} < 0.1$ mTorr, the relationship between $p_{0}^\mathrm{OMP}$ and $-\nabla n_{e}^\mathrm{sep}$ is approximately linear. As the edge density gradients grow, the neutral pressure at the wall, a proxy for the neutral flux entering the plasma, which itself balances the plasma flux transported across the edge, grows at a constant rate. Or equivalently, increases in the availability of neutrals maps directly to increases in $n_{e}^\mathrm{sep}$, $ -\nabla n_{e}^\mathrm{sep}$, and the resulting $n_{e}^\mathrm{ped}$. At some point, however, as $p_{0}^\mathrm{OMP} \rightarrow 0.1$ mTorr, this linear, diffusive-like picture breaks down coincident with the transition to the EDA H-mode. As $-\nabla n_{e} \rightarrow 200 \times 10^{20}$ m$^{-4}$, an additional mechanism, driven at high $n_{e}$ as will be discussed below, impedes further build-up of density inside the separatrix, such that $n_{e}^\mathrm{ped}$ saturates.


 As neutrals continue to accumulate in the main chamber and $p_{0}^\mathrm{OMP} \rightarrow 0.3$ mTorr, the gradient at the separatrix reduces. The separatrix density continues to increase, whereas the pedestal density does not, leading to lower density gradient, or equivalently, higher $n_{e}^\mathrm{sep}/n_{e}^\mathrm{ped}$. Without quantification of the local ionization source, it is not possible to be certain if this is purely a transport effect or also a result of lower neutral penetration.  Presumably, the degree of neutral penetration inside the pedestal, given by the neutral density decay length, $\lambda_{n_{0}}$, varies across this dataset, and changes to fueling inefficiency may indeed be at play. Even if the break in slope in $n_{e}^\mathrm{sep}$, for example, is related to opaqueness of neutrals, the flattening of the profile and subsequent saturation of $n_{e}^\mathrm{ped}$ at high $p_{0}$ is likely a transport effect and is broadly consistent with earlier analysis done using local measurements of $S_\mathrm{ion}$ using the Ly$_{\alpha}$ camera \cite{miller_enhanced_2025}. Note that similar stiffness in $n_{e}^\mathrm{ped}$ was recently observed in the EDA in a power scan on AUG \cite{gil_eda_2025}.



The ratio $n_{e}^\mathrm{sep}/n_{e}^\mathrm{ped}$, in the absence of detailed measurements of the pedestal profile, is often used as a proxy for the pedestal gradient. For ELMy discharges, the ratio $n_{e}^\mathrm{sep}/n_{e}^\mathrm{ped}$ is rather fixed, with $n_{e}^\mathrm{sep}/n_{e}^\mathrm{ped} \approx 0.4$. EDA discharges, on the other hand, can have a somewhat larger ratio, with $n_{e}^\mathrm{sep}/n_{e}^\mathrm{ped} \rightarrow 0.8$ at the highest densities (although the high $n_{e}$ EDA profile shown in dark blue in Figure \ref{fig:elmy_eda_profiles} only has $n_{e}^\mathrm{sep}/n_{e}^\mathrm{ped} \approx 0.6$). A limitation of using this ratio as a metric, rather than the more precise $\nabla n_{e}$, for example, is that it alone cannot distinguish flattening of the profile from a shift in the profile, an effect known to be present at high $n_{e}$ \cite{hughes_edge_2007, dunne_role_2017}. This is a result of the fact that the ``ped'' label moves with the profile, whereas the ``sep'' label does not (or at least it moves with the more static $T_{e}$ profile given the separatrix identification approach). Indeed, the bottom of the $n_{e}$ pedestal, for the EDA profile shown in Figure \ref{fig:elmy_eda_profiles}, is considerably lower than $n_{e}^\mathrm{sep}$, such that even though the ratio $n_{e}^\mathrm{sep}/n_{e}^\mathrm{ped}$ is high, $-\nabla n_{e}$ is as well, although the strong gradient region is now in the SOL. This distinction may have implications for performance (recalling that all plasmas in this dataset are created in the same shape); plasmas with high $n_{e}^\mathrm{sep}/n_{e}^\mathrm{ped}$, especially those in the EDA, have lower stored energy $W_\mathrm{MHD}$, a result reminiscent of that for high density EDAs from the analysis in \cite{miller_determination_2025}.

\section{The evolution of quasi-coherent fluctuations}
\label{sec:fluctuations}


Fluctuation spectra are often used to distinguish confinement regimes. It has been shown on Alcator C-Mod that in EDA H-modes, a mode called the QCM develops in the mid-frequency range of various fluctuation spectra. The QCM was routinely observed on C-Mod using a number of diagnostics \cite{greenwald_characterization_1999, lin_upgrade_1999, mazurenko_experimental_2002, terry_transport_2005}. It has been previously associated with increased particle transport \cite{hubbard_pedestal_2001, terry_transport_2005} and is thought to supply the requisite level of transport through the edge to prevent the build-up of pressure gradients triggering Type-I ELMs. The QCM has since been observed in the EDA H-mode regime achieved on a number of other devices \cite{mossessian_edge_2003, sun_experimental_2019, gil_stationary_2020, kallenbach_developments_2021, kalis_experimental_2024}, as well as in the quasi-continuous exhaust (QCE) regime on AUG \cite{faitsch_broadening_2021, faitsch_analysis_2023, dunne_quasi-continuous_2024, kalis_experimental_2024}. While this mode is thought unique to the EDA (and QCE), it is nonetheless instructive to compare the fluctuation spectra of other H-modes with that of EDA H-modes to understand how the QCM develops from possibly more broadband spectra in other regimes. It should be noted that a separate feature, called the quasi-coherent fluctuation (QCF), has been identified in the inter-ELM fluctuation spectra of ELMy H-modes on both C-Mod (including in discharges from the data set considered in the present manuscript) \cite{diallo_observation_2014, diallo_quasi-coherent_2015} and in a set of related experiments on DIII-D \cite{diallo_correlations_2015}. In those analyses, the QCF was linked to reaching the critical gradient of the kinetic ballooning mode (KBM). It was also associated with the electron temperature gradient, $\nabla T_{e}$, possibly also linked to the rebuild of the pressure profile after an ELM crash. Identification of this feature requires analysis of combined frequency-wavenumber spectra, which is out of the scope of current work. Instead, in this section, focus is placed on the evolution of the \emph{QCM} in the EDA H-mode during and after the transition from the ELMy H-mode, using a common analysis and spectral fitting routine, which is also applied to the more broadband mid-frequency inter-ELM fluctuation spectra of ELMy H-modes.

\begin{figure}
\centering
\includegraphics[width=\columnwidth]{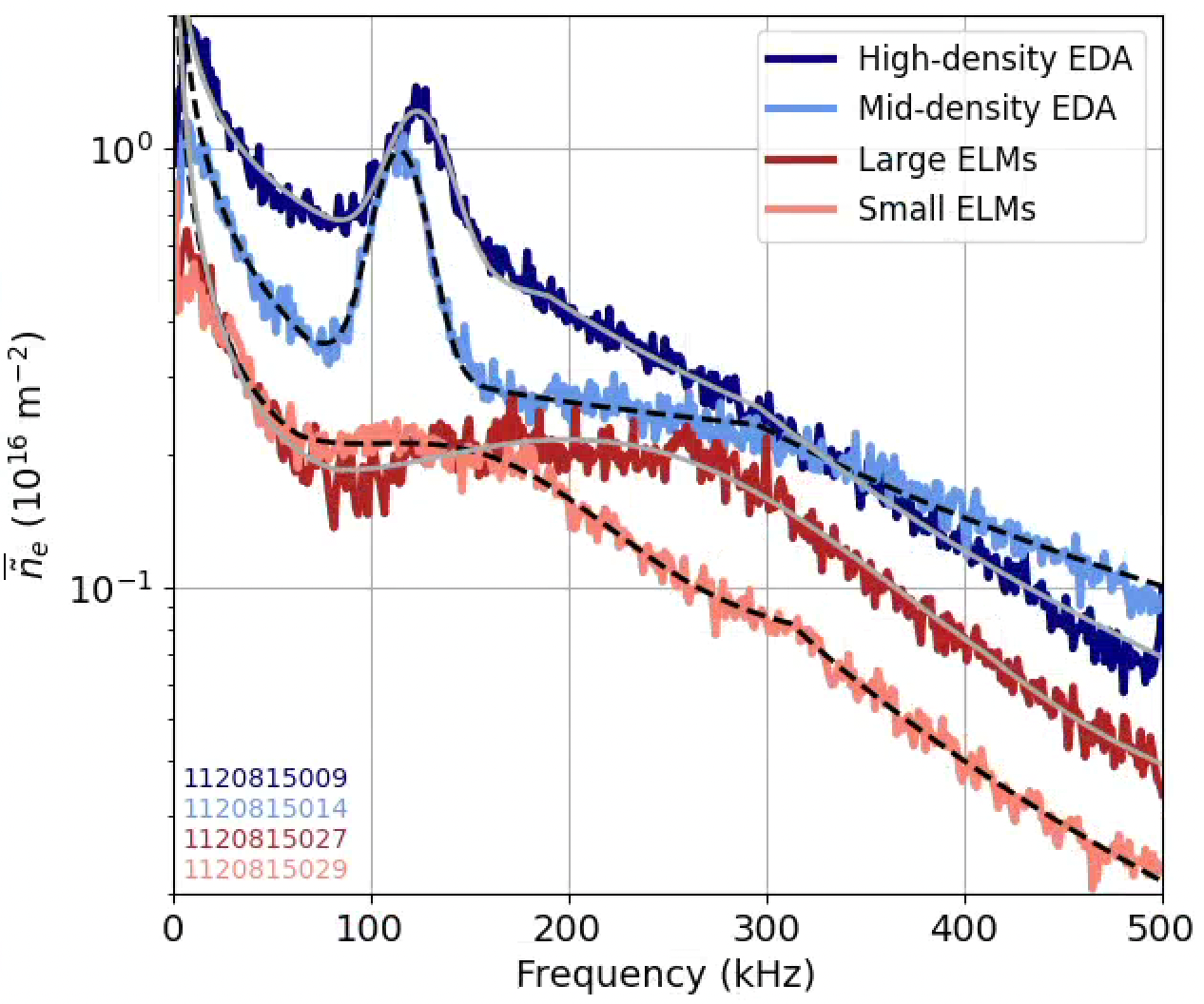}
\caption{Fourier-transformed time series of line-averaged density fluctuations as measured by the PCI for channel \#15 of four different discharges. Blue discharges are EDA H-modes and red discharges are ELMy H-modes. Solid gray and dashed black curves are used to show fits to the fluctuation spectra using Equations \ref{eq:hf} -- \ref{eq:gf}.}
\label{fig:fourier_transform}
\end{figure}



\subsection{Spectral fitting parameter dependence on edge density}
\label{sec:fitting_density}

Calculation of the characteristics of the QCM is done across regimes using measurements from the PCI diagnostic \cite{mazurenko_experimental_2002, golfinopoulos_external_2014}. The PCI on C-Mod measured fluctuations in the line-averaged density ($\tilde{\overline{n}}_{e}$) along 32 vertical chords. Though the PCI does not permit localization of the QCM in space, previous measurements of the mode using reflectometry, scanning probes, and gas puff imaging have consistently observed the mode near the separatrix \cite{terry_transport_2005, labombard_new_2014, theiler_radial_2017, golfinopoulos_edge_2018}. The analysis in \cite{diallo_quasi-coherent_2015} showed, using a similarly extensive fluctuation diagnostic suite, that the \emph{QCF} observed in the inter-ELM phase of ELMy H-modes existed within one cm of the separatrix, likely in the steep-gradient region of the pedestal. Similarities in the radial location and frequency range of the QCM and the QCF invites analysis of how the QCM evolves from the fluctuations in ELMy H-modes and how this evolution varies with local parameters in the edge.


\begin{figure*}
\centering
\includegraphics[width=1.6\columnwidth]{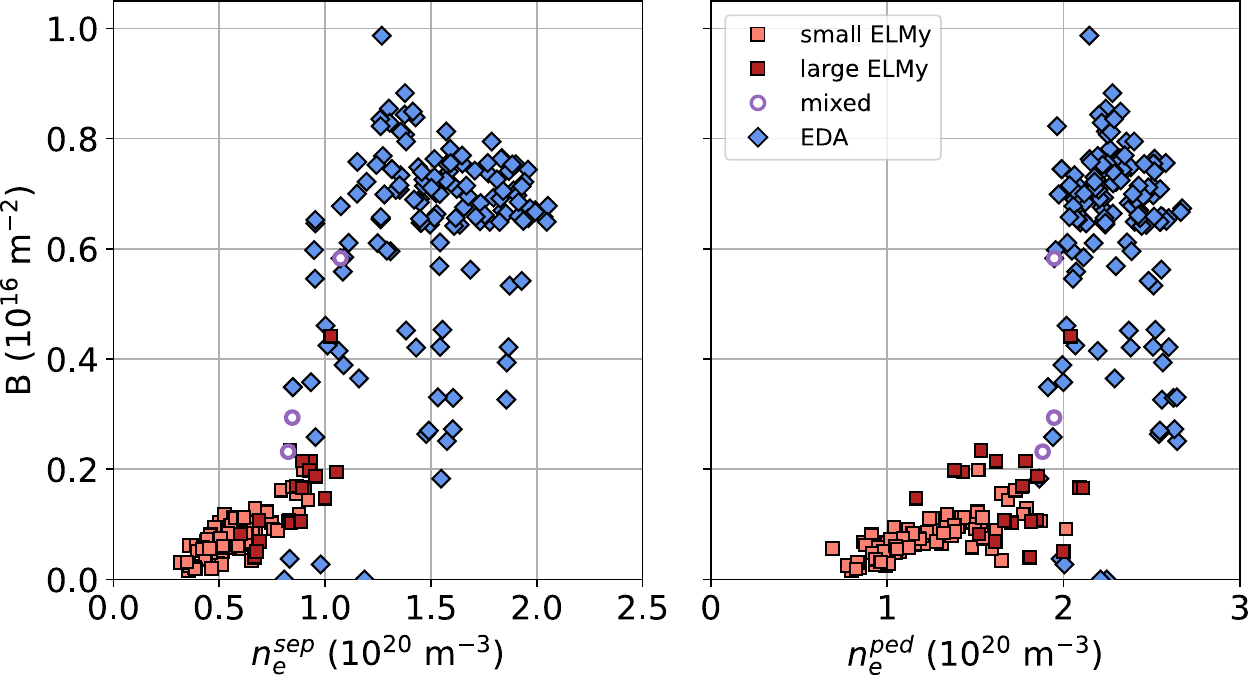}
\caption{Amplitude of Gaussian describing QC mode (top) and amplitude of background piecewise polynomials (bottom) as a function of $n_{e}^\mathrm{sep}$ (left) and as a function of $n_{e}^\mathrm{ped}$ (right), and (bottom).}
\label{fig:qcm_strength}
\end{figure*}

Figure \ref{fig:fourier_transform} shows power spectra computed using a fast Fourier transform for four representative discharges in this dataset, corresponding to the discharges whose kinetic profiles are shown in Figure \ref{fig:elmy_eda_profiles}. Differences in these spectra clearly exist, in particular, between EDA H-modes, which show a very clearly defined and narrow coherent mode in the vicinity of 100 kHz, and ELMy H-modes, whose spectra are more variable and broadband in the mid-frequency range. Following from ELM filtering analysis shown for ETS data in Section \ref{sec:exp_profs_neutrals}, spectra for ELMy H-modes are computed for the last 70\% of the ELM cycle. Generally, all of these spectra can be divided into three regions: a low-frequency region, a mid-frequency region, and a high-frequency region. It is the mid-frequency region in which the QCM typically appears for EDA discharges, as well as fluctuations possibly elevated beyond background levels for ELMy discharges. The fit function to be used follows from these observations and is inspired by that used for reflectometer analysis of density fluctuations \cite{dominguez_nodate}. The function used to fit PCI spectra across regimes designed to capture these features is a triple power law with a Gaussian bump, given by the sum:

\begin{equation}
    H(f) = P(f) + \mathcal{N}(f)   
    \label{eq:hf}
\end{equation}
where $P(f)$ is a triple piecewise power law and $\mathcal{N}(f)$ is a Gaussian, given by:

\begin{equation}
    P(f) = \begin{cases}
            A \left( \frac{f}{f_{1}} \right)^{-\alpha_{1}} & \text{if } f < f_{1},\\
            A \left( \frac{f}{f_{1}} \right)^{-\alpha_{2}}  & \text{if } f_{1} \leq f < f_{2},\\
            A \left( \frac{f_{2}}{f_{1}} \right)^{-\alpha_{2}} \left( \frac{f}{f_{2}} \right)^{-\alpha_{3}}  & \text{if } f \geq f_{2}
            \end{cases}
    \label{eq:pf}
\end{equation}

\begin{equation}
    \mathcal{N}(f) = B\,\mathrm{exp}\left( -\frac{(f-\mu)^{2}}{2\sigma^{2}} \right)
    \label{eq:gf}
\end{equation}
where $A$ and $B$ are the amplitudes of the piecewise power law and Gaussian, respectively, $f_{1}$ and $f_{2}$ are the boundaries in frequency space that separate the three regions, $\alpha_{1}$, $\alpha_{2}$, and $\alpha_{3}$ are the exponents that give the structure of the spectra in each of these regions, and $\mu$ and $\sigma$ are the center position and width of the Gaussian bump, respectively. All of these are fitting parameters, determined using a non-linear least squares optimization routine for each power spectrum. The parameters belonging to the Gaussian bump, namely $B$, $\mu$, and $\sigma$, correspond well to the size, position, and width (in frequency space) of the QCM for EDA H-modes. Note that even though there is no clear coherent mode in the ELMy H-mode spectra, the above fit function is still successful in detecting and fitting three distinct regions in frequency space. Crucially, $B$ from the Gaussian bump is thought to still adequately describe the strength of the elevated density fluctuations in the mid-frequency region. Comparison of $\mu$ and $\sigma$ for the Gaussian portion of these spectra with those from the QCM in the EDA, however is avoided, as those parameters are not considered meaningful for ELMy H-modes. For each time window, fit parameters for both $\mathcal{P}$ and $\mathcal{N}$ are determined for each of the 32 chords of the PCI diagnostic. Values reported in this section are averages of these values taken over the 32 chords.


Figure \ref{fig:qcm_strength} shows the amplitude of the Gaussian bump, $B$, as a function of $n_{e}$ at both the separatrix at left and at the pedestal top at right. From this figure, one can clearly see how the amplitude of the mode evolves from the ELMy H-mode, where fluctuations are low ($B < 0.2 \times 10^{16}$ m$^{-2}$), before becoming more coherent and considerably stronger ($B > 0.6 \times 10^{16}$ m$^{-2}$ for most EDA H-modes) as $n_{e}^\mathrm{sep} > 1.0$ m$^{-3}$ and $n_{e}^\mathrm{ped} > 2.0$ m$^{-3}$. For ELMy H-modes, $B$ appears to grow fairly linearly with the density at both locations, although there is less scatter in the figure with $n_{e}^\mathrm{sep}$ and possibly also stronger growth for large ELMy H-modes prior to the transition to the EDA. While not representative of changes to the QCM directly, changes to the mid-frequency fluctuation levels, possibly the pre-cursor for the QCM, appear to depend closely on $n_{e}$. This dependence on $n_{e}$ extends to the EDA, but only up to $n_{e}^\mathrm{sep} \approx 1.5 \times 10^{20}$ m$^{-3}$. Beyond these values of $n_{e}$, the mode amplitude saturates, at or below $0.7 \times 10^{16}$ m$^{-2}$. When plotting against $n_{e}^\mathrm{ped}$ as on the right, the lower values of the mode amplitude in the EDA ($B < 0.4 \times 10^{16}$ m$^{-2}$) appear primarily for the largest values of $n_{e}^\mathrm{ped} > 2.5 \times 10^{20}$ m$^{-3}$. It may be that these plasmas are approaching density limits, where a more pernicious (and possibly not coherent) fluctuation dominates, possibly also driving large transport. 

\subsection{Fluctuation drive from dimensionless parameters}
\label{sec:dimensionless_drive}

Finding correlations of $B$ with $n_{e}$ is useful in confirming the emergence of the QCM and robustly classifying a discharge as an EDA using a readily-measurable edge quantity. Ultimately though, the QCM is likely driven by any number of turbulent modes, so the amplitude of the QCM might also closely track some dimensionless quantity in the edge. Many studies, experimental and numerical alike, have attempted to associate the QCM with a particular instability from theory, with varying conclusions \cite{mazurenko_experimental_2002, myra_drift_2002, labombard_new_2014, chen_edge_2017, theiler_radial_2017, chen_progress_2018, kalis_experimental_2024}. More recently, a comprehensive picture of the QCM has emerged, bridging many of these observations and based on detailed fluctuation measurements across the separatrix. Using a midplane manipulator probe on AUG, it was observed that the driving location of the QCM was inside the separatrix where the radial electric field, $E_{r} < 0$ \cite{grenfell_multi-faced_2024}. At that location, it was identified to have an electromagnetic signature (as in \cite{snipes_quasi-coherent_2001}) and classified as a KBM, consistent with the observations in \cite{theiler_radial_2017}. Outside the separatrix, where $E_{r} > 0$ and $\beta_{e}$ is lower, the mode loses its EM character and becomes a drift-Alfvén wave (DAW). These measurements bridge the earlier observations on C-Mod identifying the mode as a drift wave (DW) propagating in the electron direction in the near-SOL using Langmuir probes \cite{labombard_new_2014} and as propagating in the ion diamagnetic direction (as KBMs do) near the $E_{r}$ well minimum using a gas puff imaging system \cite{theiler_radial_2017}. The transition from the DW to the KBM has been recently simulated using the GRILLIX fluid turbulence code \cite{zholobenko_fast_2026}. At the same time, the QCM has been observed at high values of $\nu^{*}$ (or $\alpha_{t}$), one of the necessary ingredients for access to the EDA. The importance of resistivity is captured in a theory suggesting that the QCM results from the coupling of the aforementioned DAW with the RBM \cite{myra_drift_2002, chen_edge_2017, chen_progress_2018}. This strong coupling is enabled at high $\beta$ \cite{tang_destabilization_1976}, requiring also high $\alpha_{t}$ to destabilize the RBM. Together, these provide the EM and resistive drive thought to be important for emergence of the QCM and transition to the EDA.

\begin{figure}
\centering
\includegraphics[width=0.8\columnwidth]{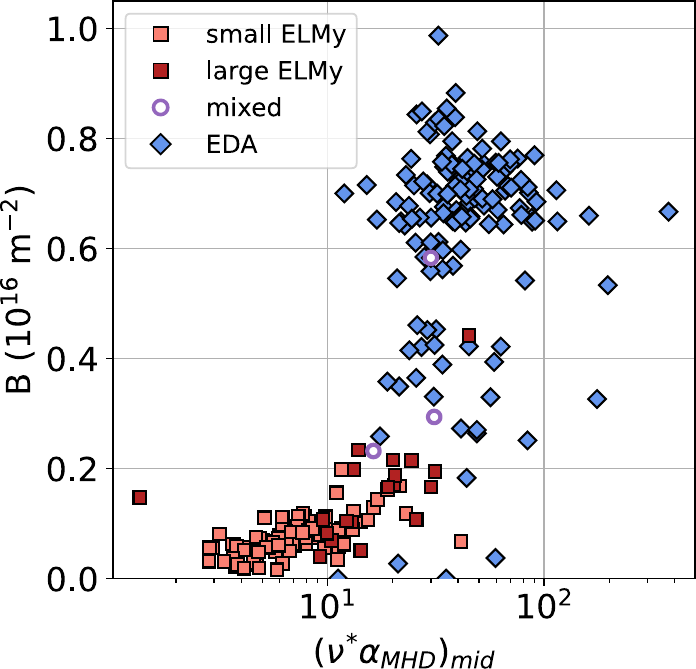}
\caption{Amplitude of Gaussian describing QC mode against product of $\nu^{*}$ and $\alpha_\mathrm{MHD}$ evaluated at mid-pedestal.}
\label{fig:nustar_alphamhd}
\end{figure}

The importance of both high collisionality and high edge plasma pressure in driving the QCM and accessing the EDA H-mode has also been identified experimentally for EDAs on Alcator C-Mod \cite{hughes_observations_2002, hughes_edge_2007}. In those works, EDAs were observed to exist at higher $\nu^{*}$ and normalized pressure gradient, $\alpha_\mathrm{MHD}$, than the so-called ``ELM-free'' H-mode, another regime free of Type-I ELMs routinely achieved on C-Mod but inherently transient as a result of its high impurity content. Fig. 1 of \cite{hughes_edge_2007} shows that if the product, $\nu^{*}\alpha_\mathrm{MHD}$, computed at the center of the $p_{e}$ pedestal, i.e. the location of maximum pressure gradient, is high enough, a discharge will be an EDA. To make direct contact with this work, Figure \ref{fig:nustar_alphamhd} plots $B$ against the product of these parameters, $(\nu^{*}\alpha_\mathrm{MHD})_\mathrm{mid}$, computed also at the middle of the $p_{e}$ pedestal. Note that as in that work, $\alpha_\mathrm{MHD}$ is computed using the approximation of a circular cross-section, i.e. $\alpha_\mathrm{MHD} = \frac{2\mu_{0}q_{95}^{2}R_{0}}{B_{T}^{2}}\nabla p$, instead of also accounting for the flux surface geometry, as is done in Section \ref{sec:density_model_validation}. $B$ trends similarly with $(\nu^{*}\alpha_\mathrm{MHD})_\mathrm{mid}$ as it does with $n_{e}$ -- it increases slowly in the ELMy H-mode, quickly at the transition value, and then saturates. The large growth of the QCM and transition to the EDA H-mode occurs at a critical value of $(\nu^{*}\alpha_\mathrm{MHD})_\mathrm{mid} \sim 20-30$, consistent with the range of values in the product of these parameters in EDAs identified earlier analysis.

\begin{figure*}
\centering
\includegraphics[width=1.9\columnwidth]{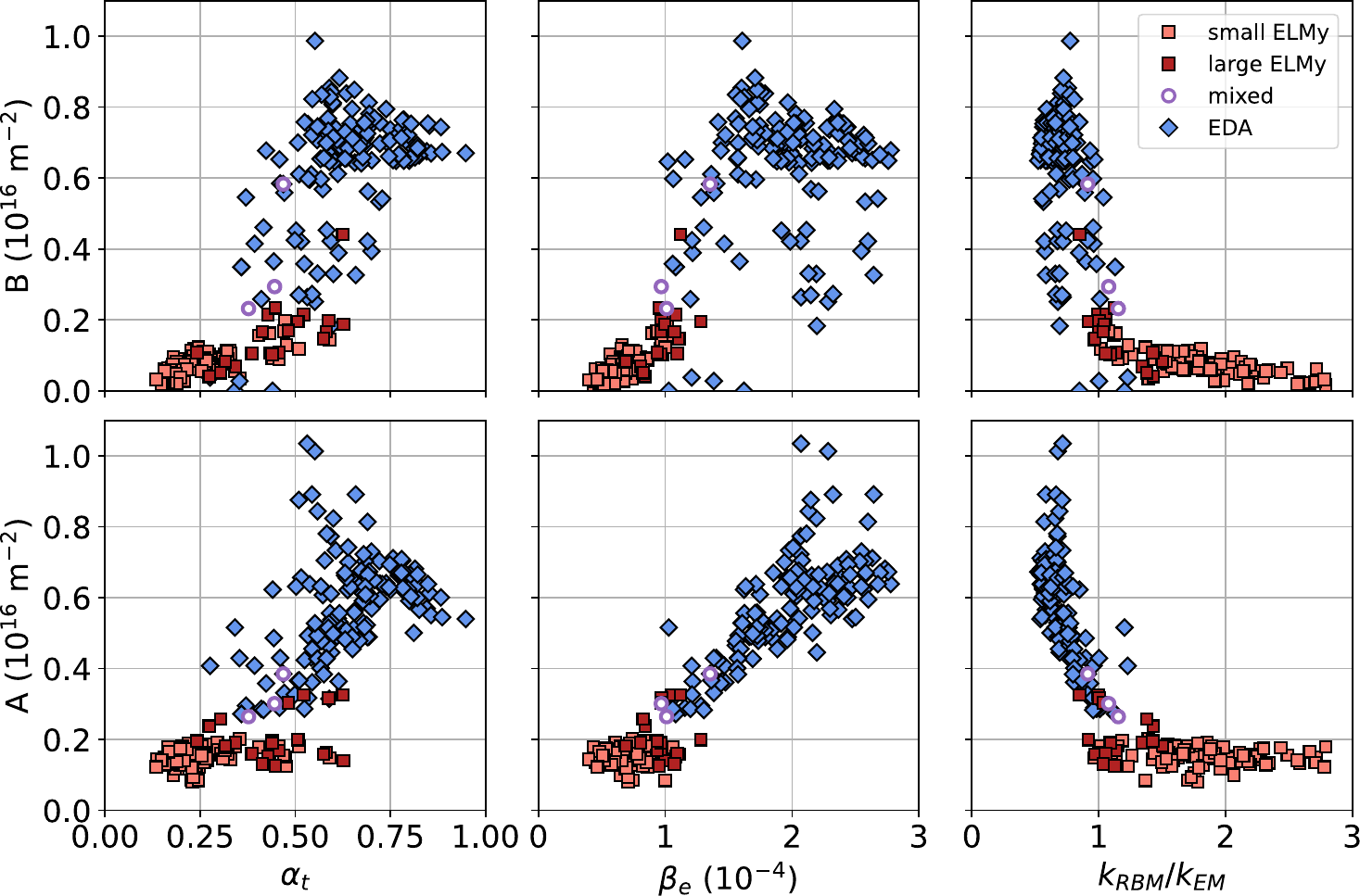}
\caption{Amplitude of Gaussian describing QCM as a function of $\alpha_{t}$ (left), $\beta_{e}$ (center), and $k_\mathrm{RBM}/k_\mathrm{EM}$ (right).}
\label{fig:qcm_transition_params}
\end{figure*}

Previous work on C-Mod \cite{labombard_new_2014, theiler_radial_2017, golfinopoulos_edge_2018} and AUG \cite{grenfell_multi-faced_2024, kalis_experimental_2024} has suggested that the QCM may be localized close to the separatrix (which for many C-Mod H-modes is close to the center of the pedestal). Additionally, recent efforts to locate the separatrix on both AUG and C-Mod have shown success in using dimensionless separatrix parameters from the SepOS framework to describe turbulent transport and the resulting edge operational space \cite{eich_turbulence_2020, eich_separatrix_2021, miller_determination_2025, miller_fluxes_2025}. To connect with these observations, one final set of figures is produced, Figure \ref{fig:qcm_transition_params}, but now as a function of three different local parameters identified in \cite{miller_determination_2025} to describe the transition between ELMy and EDA H-modes. The top row shows the same mode amplitude, $B$, as plotted in Figures \ref{fig:qcm_strength} and \ref{fig:nustar_alphamhd}. The first parameter plotted at left is $\alpha_{t}$, the normalized collisional drive parameter introduced in Section \ref{sec:exp_profs_neutrals}. In the center plot, $B$ is plotted against $\beta_{e}$, also at the separatrix. Recall that $\alpha_{t}$ and $\beta_{e}$ are functionally equivalent to $\nu^{*}$ and $p_{e}$, respectively, given relatively constant edge $q$ and $B_{t}$ in this dataset. Finally, the plot at right uses the ratio of two wavenumbers from the SepOS model on the abscissa: that of the characteristic wavenumber for fluctuations from resistive ballooning mode (RBM) turbulence, $k_\mathrm{RBM} = \sqrt{\frac{\alpha_{c}\sqrt{\omega_{B}}}{\alpha_{t}}}$ to that for electromagnetic (EM) turbulence, $k_\mathrm{EM} = \sqrt{\frac{\beta_{e}}{\mu}}$. In these expressions, $\alpha_{c}$ is the normalized critical pressure gradient, $\omega_{B} = \frac{2\lambda_{p_{e}}}{R_\mathrm{geo}}$ gives the curvature drive, with $\lambda_{p_{e}}$ the gradient scale length of $p_{e}$, and $\mu$ is the reduced mass. All quantities on the abscissas of Figure \ref{fig:qcm_transition_params} are evaluated at the separatrix. Details and derivation of these wavenumbers can be found in \cite{eich_separatrix_2021}. In that work, the ratio $k_\mathrm{RBM}/k_\mathrm{EM}$ was used to describe the transition to the L-mode density limit and in \cite{miller_determination_2025}, it was suggested that these wavenumbers, evaluated using H-mode parameters, also contained valuable information for describing the transition ELMy-EDA H-mode transition. 

In general, Figure \ref{fig:qcm_transition_params} does not unequivocally point to one parameter as being most important for describing the evolution of the QCM, but it does allow additional observations. First, in the transition from ELMy to EDA discharges, $B$ increases with both $\alpha_{t}$ and $\beta_{e}$. This happens most strongly as critical values in these two parameters corresponding to the ELMy-EDA transition are reached, i.e. $\alpha_{t} = 0.55$ and $\beta_{e} = 10^{-4}$. At this transition, however, there is somewhat more scatter in $B$, with the largest scatter against $\alpha_{t}$. Regardless, the data in the top left and center panels of Figure \ref{fig:qcm_transition_params} are generally consistent with the picture of the QCM requiring both resistive and EM drive, in that large values of $B$ are only found at high $\alpha_{t}$ and $\beta_{e}$. Furthermore, a threshold in $\beta_{e}$ is consistent with recent linear local GENE simulations showing sensitivity of the QCM growth rate with $\beta_{e}$. The top right panel shows perhaps the clearest trend -- $B$ decreases with $k_\mathrm{RBM}/k_\mathrm{EM}$, more strongly for EDAs with values less than unity, and slightly less so for ELMy H-modes with values greater than unity. As with the other two parameters, the largest change in $B$ occurs as $k_\mathrm{RBM}/k_\mathrm{EM} \rightarrow 1$, the criterion identified to describe the transition to the EDA in \cite{miller_determination_2025}. As $k_\mathrm{RBM} \sim k_\mathrm{EM}$ and large ELMs are supplanted by a clearly coherent mode, the strength of the QCM increases quickly. The observation that the strength of the fluctuation depends on both the wavenumbers of RBM and EM turbulence provides further evidence that interplay of these two types of turbulence (and not one alone) contribute to the strength of the fluctuations from this mode. This rightmost plot also provides the connection to the DAW-RBM coupling picture for emergence of the QCM, suggesting that that this coupling coincides with the fluctuations from the RBM becoming EM ($k_\mathrm{RBM} \sim k_\mathrm{EM}$). Finally, the observation of saturation and weakening of the QCM at high $n_{e}$ made above can also be gauged from these plots. The saturation especially with $\alpha_{t}$ is  reminiscent of observations of the dependence of fluctuation levels on $\alpha_{t}$ made on both AUG and EAST \cite{eich_turbulence_2020, li_study_2025}.

The bottom row of Figure \ref{fig:qcm_transition_params} shows another parameter from the spectral fit function -- the amplitude, $A$, of the piecewise polynomial, $\mathcal{P}$. Unlike $B$, which is exactly the height of the Gaussian (ignoring the background), $A$ is not as readily interpretable from the power spectrum directly. Regardless, under the reasonable assumption that $\sigma \ll \mu$ for the Gaussian, $\mathcal{N}(f)$, one can approximate $\mathcal{N}(f=0)$ as small and take $H(f=0) = P(f=0) +  \mathcal{N}(f=0) \approx P(f=0) = A$. In this way, the fit parameter $A$ is representative of the background level of fluctuations, $\overline{\tilde{n}}_{e}$ near $f=0$. It is important to note that since the PCI chords also pass through the core, this is not strictly an edge fluctuation level, but rather one of the plasma as a whole. It is not unreasonable to expect, however, that $n_{e}$ fluctuations are strongest near the edge. As such, it is still instructive to compare the strength of the background, approximately $A$, against turbulence drive parameters evaluated at the separatrix, as is done in the bottom row of Figure \ref{fig:qcm_transition_params}. In the ELMy phase, $A$ does not vary much as a function of any of the parameters, staying below $0.2 \times 10^{16}$ m$^{-2}$, except for a small subset of large ELMy H-modes. When in the EDA, however, $A$ increases immediately, trending almost linearly with both $\alpha_{t}$ and $\beta_{e}$. Interestingly, at $\alpha_{t} > 0.75$, $A$ rolls over and begins to decrease with $\alpha_{t}$. The same does not happen with $\beta_{e}$, with a fairly linear trend throughout the EDA dataset. As above, this trend is clearest in the rightmost plot against $k_\mathrm{RBM}/k_\mathrm{EM}$ -- in the ELMy phase ($k_\mathrm{RBM} > k_\mathrm{EM}$), there is no change in $A$. At the point where $k_\mathrm{RBM} \sim k_\mathrm{EM}$, however, $A$ jumps, and begins increasing stronger than linearly as $k_\mathrm{RBM}/k_\mathrm{EM}$ decreases.

These trends are consistent with the observation of higher $D_{\alpha}$ activity in the high-$n_{e}$ EDA compared to the mid-$n_{e}$ EDA H-mode in Figure \ref{fig:dalpha_traces} (and certainly than that in the ELMy H-mode). If $A$ is indeed representative of the background level of turbulence, much of which in the edge is dominated by filamentary transport, it is not unreasonable to expect that $A$ might be a proxy for the level of such transport. As these parameters mediating collisional and EM drive increase, so does the background turbulence and possibly the level of edge filamentary transport. These trends are also reminiscent of observations of filament activity of EDAs on both C-Mod \cite{labombard_new_2014} and AUG \cite{gil_eda_2025}. Furthermore, a link may exist between the increase of $A$ with these parameters and the filaments of the QCE \cite{faitsch_broadening_2021}. Finally, it is worthwhile to note that the background fluctuation level continues to grow, \emph{despite} saturation or a possible rollover in the QCM height. At large values of these dimensionless parameters, it may not only be the QCM providing strong transport, but possibly also another instability responsible for high background fluctuation levels.

\section{Pedestal density modeling and validation}
\label{sec:density_model_validation}

The following two sections show the results of two different computational tools used to understand the pedestal structure of current devices and eventually predict it on future ones. The current section begins with validation of the recently-developed Saarelma-Connor model for the density pedestal of H-mode discharges \cite{saarelma_testing_2023}. Validation of this model on Alcator C-Mod extends its validation to densities up to $3 \times 10^{20}$ m$^{-3}$, up to three times higher than that of previously-tested devices. Further, the model has, up to now, only been tested for Type-I ELMy H-modes, which are presumed to be limited by different instabilities in the edge than non-ELMing H-modes. This validation thus extends the model also to the non-ELMing, EDA regime \cite{saarelma_density_2024}.


\subsection{Validation of Saarelma-Connor density prediction model on Alcator C-Mod}
\label{subsec:sc_model}

The Saarelma-Connor density prediction model combines a reduced model for pedestal plasma transport with one for neutral penetration. It was first tested on JET data \cite{saarelma_testing_2023} and now also on MAST-U and AUG \cite{saarelma_density_2024}. Work to extend the validation to DIII-D as well is underway and appears promising \cite{dunsmore_validating_2025}. The density prediction model can predict $n_{e}^\mathrm{ped}$ using two different modes. In the ``standalone'' mode, the model takes in a user-prescribed $T_{e}$ pedestal profile and predicts an $n_{e}$ pedestal profile. This yields $n_{e}^\mathrm{ped}$, using other experimental engineering parameters like the input power, the magnetic fields and plasma current, the plasma shape, etc., in addition to model settings, as described below. In the ``coupled'' mode, the model circumvents the need to prescribe a $T_{e}$ profile by invoking the Europed \cite{saarelma_self-consistent_2019} (or EPED \cite{snyder_development_2009}) constraint that the pedestal pressure profile is near stability boundaries for both KBMs and peeling-ballooning modes (PBMs). The standalone prediction is first used to generate a set of $n_{e}$ and $T_{e}$ pedestal profiles that satisfy the combined neutral and plasma transport model, and these are then fed into the Europed model to select the profile marginally stable to PBMs. The initial validation of the density prediction model presented here is limited to the standalone prediction for now, but usage of the ``coupled'' version is the subject of ongoing work. Section \ref{sec:eped} will motivate this effort with comparison to EPED simulations.

The neutrals portion of this model is adapted from models tested on DIII-D by Groebner and Mahdavi \cite{groebner_role_2002, mahdavi_physics_2003} for pedestal build-up driven by neutral ionization and charge exchange (CX) in the pedestal. The Saarelma-Connor model adapts the Groebner-Mahdavi equations into a two-fluid  neutral transport model described by the following three equations:

\begin{equation}
    \nabla \cdot (D_\mathrm{ped} \nabla n_{e}) = -n_{e}(n_\mathrm{FC} + n_\mathrm{CX})S_{i}
    \label{eq:continuity}
\end{equation}

\begin{equation}
    \nabla \cdot (V_\mathrm{FC}n_\mathrm{FC}) = -n_{e}(n_\mathrm{FC}S_{i} + n_\mathrm{CX}S_\mathrm{CX})
    \label{eq:fc}
\end{equation}

\begin{equation}
    \nabla \cdot (V_\mathrm{CX}n_\mathrm{CX}) = -n_{e}(n_\mathrm{CX}S_{i} - \frac{1}{2}n_\mathrm{FC}S_\mathrm{CX})
    \label{eq:cx}
\end{equation}
where $D_\mathrm{ped}$ is the pedestal particle transport coefficient, $S_{i}$ and $S_\mathrm{CX}$ are ionization and CX rates, respectively. Velocities, $V_\mathrm{FC}$ and $V_\mathrm{CX}$, and densities, $n_\mathrm{FC}$ and $n_\mathrm{CX}$, are Franck-Condon (FC) and CX velocities and densities, respectively. Expressions for $S_{i}$ and $S_\mathrm{CX}$ are given by $S_{i} = \sigma_{i}v_{\mathrm{th},e}$ and $S_\mathrm{CX} = \sigma_\mathrm{CX}v_{\mathrm{th},i}$, with $\sigma_{i}$ and $\sigma_\mathrm{CX}$ the ionization and CX cross-sections, respectively, and $v_{\mathrm{th},e}$ and $v_{\mathrm{th},i}$ the electron and ion thermal velocities, respectively. FC neutrals are considered born from molecular dissociation and have a radial velocity given by $|V_{\mathrm{FC},r}| = \sqrt{8E_\mathrm{FC}/\pi^{2}m_{i}}$, where $E_\mathrm{FC} \sim 3$ eV. The radial velocity of CX neutrals depends on the ion temperature, $T_{i}$, according to $|V_{\mathrm{CX,r}}| = \sqrt{2T_{i}/\pi m_{i}}$, with $m_{i}$, the ion mass. Equations \ref{eq:continuity} -- \ref{eq:cx} are simply the continuity equations for the three primary types of particles present in the pedestal: electrons, FC neutrals, and CX neutrals. Their solution gives the steady-state density of neutrals and electrons. The factor of $\frac{1}{2}$ on the final term of Equation \ref{eq:cx} accounts for the assumption that the outward flux of fast CX neutrals are lost. With an appropriate choice for $D_\mathrm{ped}$, this model can be solved readily.

\begin{figure*}
\centering
\includegraphics[width=1.6\columnwidth]{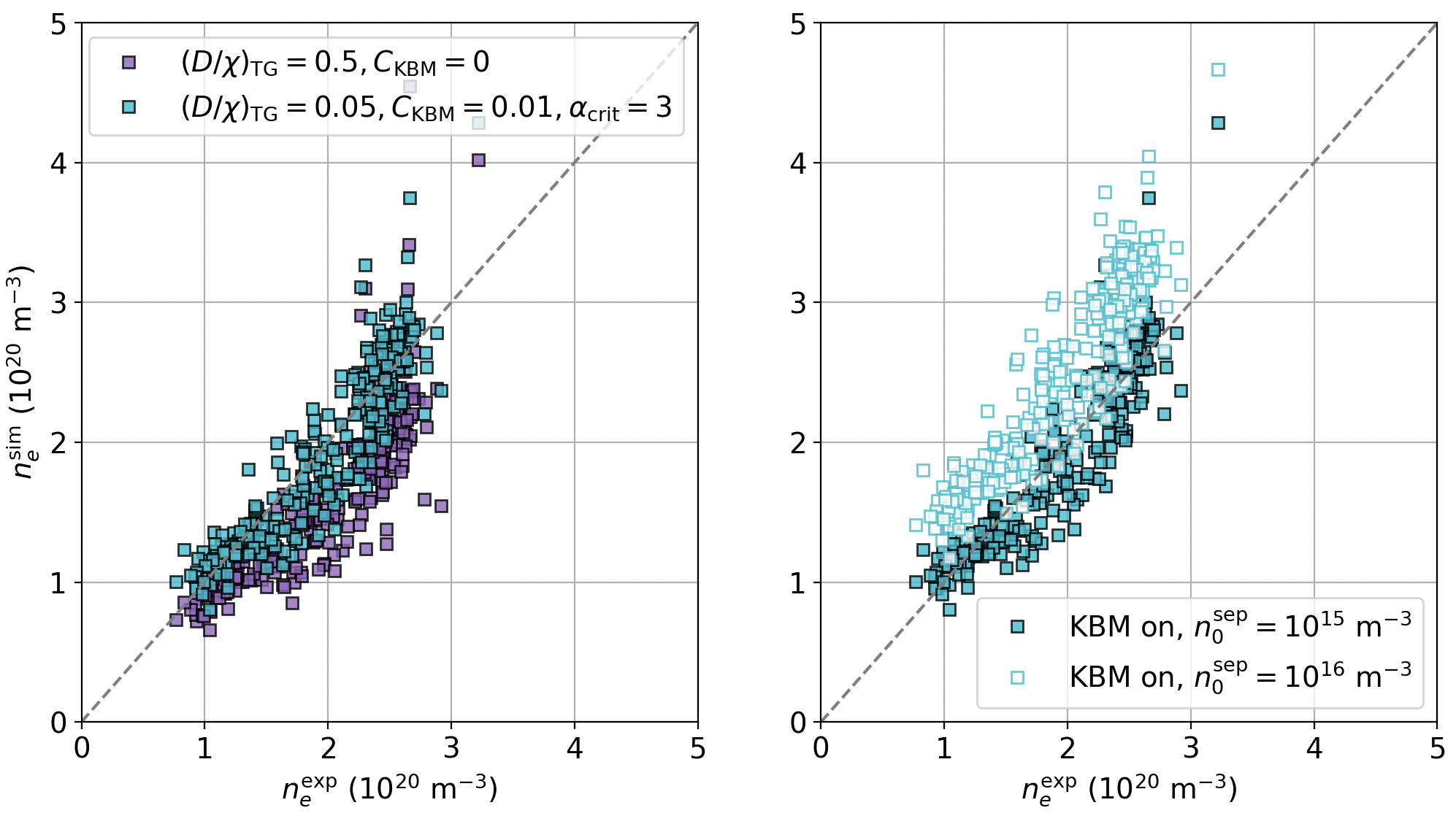}
\caption{Results of pedestal density prediction for ELMy-EDA dataset with different transport settings (left) and different $n_{0}$ boundary conditions (right). Closed purple squares show $\left( \frac{D}{\chi} \right)_\mathrm{TG} = 0.5$, $C_\mathrm{KBM} = 0$, and $n_{0}^\mathrm{sep} = 10^{15}$ m$^{-3}$. Closed turquoise squares show $\left( \frac{D}{\chi} \right)_\mathrm{TG} = 0.05$, $C_\mathrm{KBM} = 0.01$, $\alpha_\mathrm{crit} = 3$, and $n_{0}^\mathrm{sep} = 10^{15}$ m$^{-3}$. Open turquoise squares show $\left( \frac{D}{\chi} \right)_\mathrm{TG} = 0.05$, $C_\mathrm{KBM} = 0.01$, $\alpha_\mathrm{crit} = 3$, and $n_{0}^\mathrm{sep} = 10^{16}$ m$^{-3}$.}
\label{fig:results_elmy_eda_transport}
\end{figure*}

The model assumes that $D_\mathrm{ped}$ depends on the sum of diffusion from these transport mechanisms, neoclassical and turbulent alike, as follows:

\begin{equation}
    D_\mathrm{ped} = D_\mathrm{neo} + D_\mathrm{KBM} + D_\mathrm{TG}
    \label{eq:d_ped}
\end{equation}
where $D_\mathrm{neo}$, $D_\mathrm{KBM}$, and $D_\mathrm{TG}$ are transport coefficients corresponding to neoclassical transport, KBM-driven transport, and transport from modes driven by a thermal gradient (TG), according to the following:

\begin{equation}
    D_\mathrm{neo} = \frac{\chi_\mathrm{neo}}{2} = 0.05\left( \frac{\rho_{s}^{2}c_{s}}{a} \right)
    \label{eq:d_neo}
\end{equation}

\begin{equation}
    D_\mathrm{KBM} = 
    \begin{cases}
        C_\mathrm{KBM}(\alpha - \alpha_\mathrm{crit})\left( \frac{\rho_{s}^{2}c_{s}}{a} \right), & \alpha \geq \alpha_{crit} \\
        0, & \alpha < \alpha_{crit}
    \end{cases}
    \label{eq:d_kbm}
\end{equation}

\begin{equation}
    D_\mathrm{TG} = \left( \frac{D}{\chi} \right)_\mathrm{TG} \frac{P_\mathrm{tot,e}}{Sn_{e}\nabla T}
    \label{eq:d_tg}
\end{equation}
where $\frac{P_\mathrm{tot,e}}{Sn_{e}\nabla T} = \chi_{e,\mathrm{eff}}$, the effective thermal diffusivity, with $P_\mathrm{tot,e}/S$, the total power flux carried by the electrons through the pedestal, and $\alpha$, the normalized pressure gradient including geometric factors, given by:

\begin{equation}
    \alpha = \frac{2dV/d\psi}{(2\pi)^{2}}\left(\frac{V}{2\pi^{2}R_{0}}\right)^{1/2} \mu_{0} \frac{dp}{d\psi}
    \label{eq:alpha}
\end{equation}
where $V$ is the plasma volume and $\psi$ is the poloidal magnetic flux. Note that the TG-driven particle transport coefficient, $D_\mathrm{TG}$, is computed from the $\chi_{e,\mathrm{eff}}$ coefficient (and the specified $D/\chi$ ratio).

Although these are reduced equations for plasma transport, which typically requires calculation using a sophisticated neoclassical or gyrokinetic transport code, Equations \ref{eq:d_neo} -- \ref{eq:d_tg} together provide an estimate for how particle transport might depend on local pedestal plasma profile parameters. The turbulent transport is given by the combination of the KBM and any TG-driven modes, including both ion TG and electron TG modes. Despite arising from physical intuition for the drive of these various modes from a number of higher-fidelity simulations, the expressions for the various transport channels presented here still include free parameters. Specifically, $\left( \frac{D}{\chi} \right)_\mathrm{TG}$, $C_\mathrm{KBM}$, and $\alpha_\mathrm{crit}$ are user-supplied. The former expresses the amount of particle transport relative to thermal transport carried by TG modes, and the latter two are related to the KBM, including the magnitude of the diffusion, as well as the critical gradient at which this mode onsets. Note that the critical gradient used here, $\alpha_\mathrm{crit}$, is that for ideal MHD modes, and it roughly differs from that for the KBM, $\alpha_\mathrm{crit}^\mathrm{KBM}$, by a factor of $\frac{1}{1 + a/R}$ \cite{chu_kinetic_1978}, i.e. $\alpha_\mathrm{crit}^\mathrm{KBM} = \frac{\alpha_\mathrm{crit}}{1 + a/R}$. Finally, the density prediction model accepts $n_{e}^\mathrm{sep}$ and $n_{0}^\mathrm{sep}$ as boundary conditions.



Figure \ref{fig:results_elmy_eda_transport} shows initial results of the standalone prediction of the model for this dataset. The left panel in this figure shows the results for different choices of $\left( \frac{D}{\chi} \right)_\mathrm{TG}$, $C_\mathrm{KBM}$, and $\alpha_\mathrm{crit}$. The choice of $\left( \frac{D}{\chi} \right)_\mathrm{TG} = 0.5$, and $C_\mathrm{KBM} = 0$ is shown with purple squares. It corresponds to no KBM transport and represents adjusting the ratio of particle to thermal transport driven by TG modes to compensate for the lack of KBM transport, as was done in  \cite{saarelma_density_2024}. These choices were found to facilitate the use of the coupled model (which already includes a constraint for the KBM based on stability) on that multi-device validation. It is used here as an initial comparison for the standalone model for C-Mod. These choices underpredict $n_{e}^\mathrm{ped}$ for most discharges across the scan, presumably because TG-driven transport is too high. Lowering $\left( \frac{D}{\chi} \right)_\mathrm{TG}$ to 0.05, adding weak KBM-driven transport with $C_\mathrm{KBM} = 0.01$, and using a slightly higher value of $\alpha_\mathrm{crit} = 3$ than the value of $\alpha_\mathrm{crit} = 2$ found most suitable for other devices gives the results shown with turquoise squares. Full calculation of $\alpha_\mathrm{crit}$ for the KBM is out of the scope of this work. A rough estimate for $\alpha_{c}$, albeit at the separatrix, can be estimated using the formula, $\alpha_{c} = \kappa_\mathrm{geo}^{1.2} (1+ 1.5\delta)$ used in \cite{eich_separatrix_2021} and derived from ideal stability simulations. This yields an average value of $\alpha_{c} = 2.6 \pm 0.03$ across this dataset, which motivates the comparison with a slightly higher value of $\alpha_\mathrm{crit} = 3$ for KBM onset in the density prediction model. This setting represents changing from purely TG-driven transport to replacing some TG-driven with some KBM-driven transport. As such, the biggest impact is expected where pressure gradients (and thus $\alpha$) are high, and especially where, in addition, temperature gradients are lower. These different transport settings improve the prediction across most of the dataset, but have least impact at lowest and highest $n_{e}$, where pressure gradients are generally lowest. Indeed, highest sensitivity to this change and most improvement in model agreement is found for large-ELM discharges, which have large $\alpha_\mathrm{MHD}$ due primarily to large $\nabla T$, and for moderate-$n_{e}$ EDAs, which also have large $\alpha_\mathrm{MHD}$ although due instead to large $\nabla n$. 

\begin{figure*}
\centering
\includegraphics[width=1.6\columnwidth]{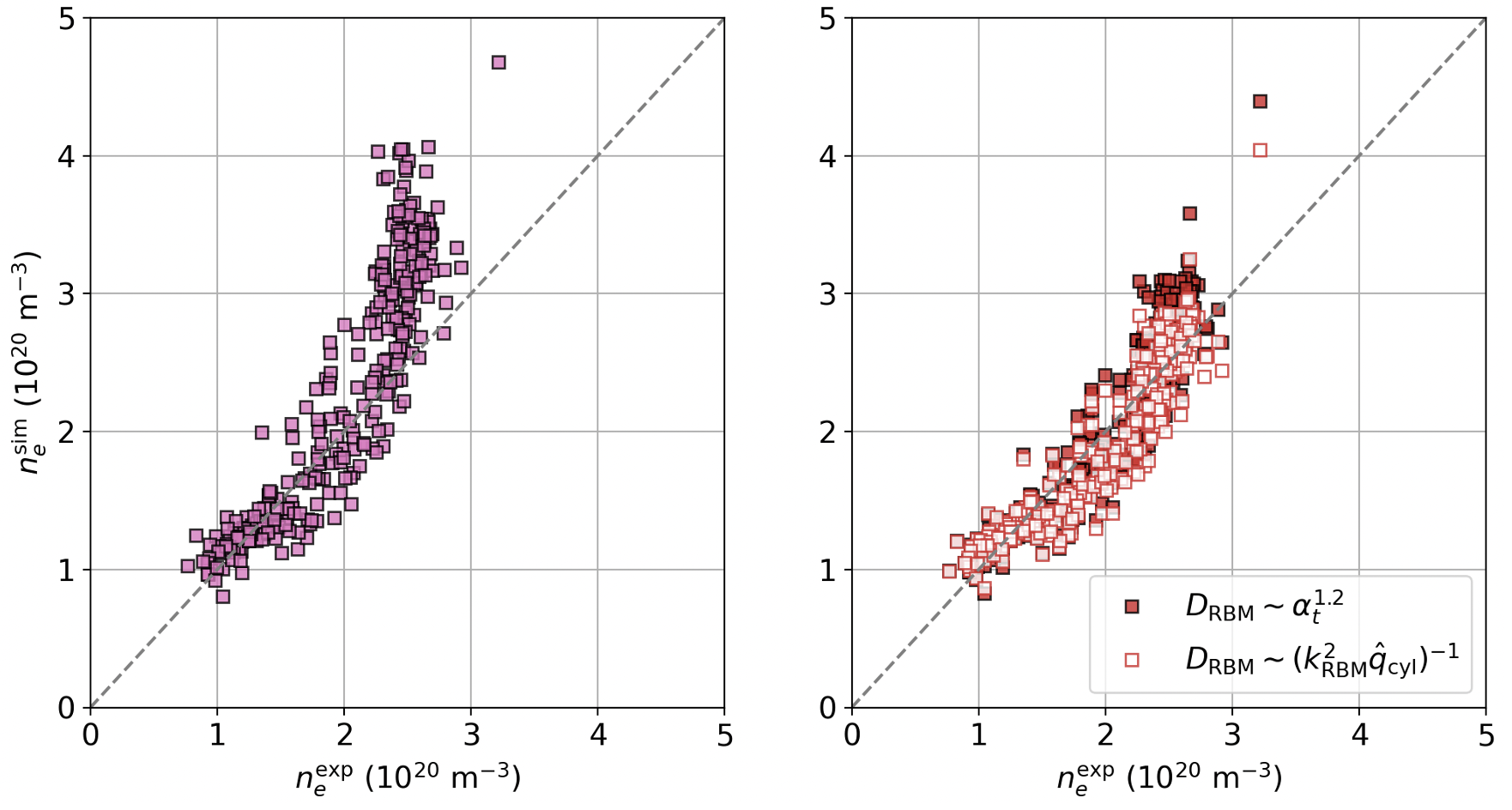}
\caption{Results of pedestal density prediction for ELMy-EDA dataset, using same settings as for solid turquoise squares from Figure \ref{fig:results_elmy_eda_transport}, including dependence of $n_{0}^\mathrm{sep}$ on experimental $p_{0}^\mathrm{OMP}$ (left) and adding this and the effect of RBM transport channel (right). At right, solid red squares use Equation \ref{eq:d_rbm_alphat} with $C_\mathrm{RBM}$ = 0.039 and open red squares use Equation \ref{eq:d_rbm_krbm} with $C^{k}_\mathrm{RBM}$ = 0.086.}
\label{fig:results_elmy_eda_include_neutrals_rbm}
\end{figure*}

All predictions in the left panel of Figure \ref{fig:results_elmy_eda_transport} use the same boundary condition for neutrals as was used in the prediction for other devices: $n_{0}^\mathrm{sep} = 10^{15}$ m$^{-3}$. Comparison with previously-published values of $n_{0}^\mathrm{sep}$ on C-Mod \cite{miller_particle_2025} indicates that this may be an underestimate for C-Mod plasmas, for which $n_{0}^\mathrm{sep}$ has been reported on the order of $10^{16}$ m$^{-3}$ and can reach up to $10^{17}$ m$^{-3}$, at least for EDA H-modes in the the standard C-Mod shape. Given the repositioning of the outer strike point in the plasma shape used for this dataset and the lower-density regimes accessed, it is possible that $n_{0}^\mathrm{sep}$ differs from that in the conventional shape, but it is difficult to know by how much. Furthermore, given large changes to $p_{0}^\mathrm{OMP}$ in this dataset, it is not unreasonable to assume that $n_{0}^\mathrm{sep}$ might itself vary significantly within the dataset. In the absence of a measurement of $n_{0}^\mathrm{sep}$ to calibrate this boundary condition, the right panel of Figure \ref{fig:results_elmy_eda_transport} shows the impact of changing $n_{0}^\mathrm{sep}$ from $10^{15}$ m$^{-3}$ to $10^{16}$ m$^{-3}$ uniformly across the dataset. 

Two things are apparent from this exercise. First, the prediction is generally better when using the $10^{15}$ m$^{-3}$ boundary condition. Second, the change in $n_{e}^\mathrm{sim}$ engendered by a change in $n_{0}^\mathrm{sep}$ is slightly different at low $n_{e}^\mathrm{exp}$ and at high $n_{e}^\mathrm{exp}$. At $n_{e}^\mathrm{exp} = 1 \times 10^{20}$ m$^{-3}$, $n_{e}^\mathrm{sim}$ increases by about 50\% when changing the neutral density boundary condition. At $n_{e}^\mathrm{exp} = 2.5 \times 10^{20}$ m$^{-3}$, the change in $n_{e}^\mathrm{sim}$ is no more than 20\% on average. There is more scatter in the percent change here, however, since this portion of the dataset has a larger density of data, as this is where EDAs, at relatively fixed $n_{e}^\mathrm{exp}$, live. The observed difference in the effect of modifying the neutral boundary condition at low and high $n_{e}$ may be explained by neutral penetration. At low $n_{e}$, $\lambda_{n_{0}}$ is large such that a change to the source of neutrals can propagate into the pedestal more easily than at high $n_{e}$, where $\lambda_{n_{0}}$ is low, and $n_{e}$ may be set instead by transport.


\subsection{Estimation of $n_{0}^\mathrm{sep}$ and addition of RBM particle transport at high $n_{e}$}
\label{subsec:n0_rbm_transport}


In order to increase confidence in an appropriate $n_{0}^\mathrm{sep}$ boundary condition and assess variation in this value throughout the dataset, kinetic neutral simulations of the main chamber SOL plasma are performed using KN1D \cite{labombard_kn1d_nodate}. The code is used to characterize $n_{0}^\mathrm{sep}$ across this dataset by parameterizing it in terms of the experimentally-available $p_{0}^\mathrm{OMP}$. This is done by running 10 KN1D simulations using characteristic plasma profiles at 10 different logarithmically-spaced values of $p_{0}^\mathrm{OMP}$ that span this dataset. Details of the procedure can be found in \ref{sec:kn1d_sims}. The resulting fit is given by $n_{0}^\mathrm{sep}\mathrm{[10^{15} m^{-3}]} = 38.5 p_{0}^\mathrm{OMP} \mathrm{[mTorr]}$, which is then used to estimate $n_{0}^\mathrm{sep}$ for the Saarelma-Connor model. The results for the parameterized neutral boundary condition are shown in the left panel of Figure \ref{fig:results_elmy_eda_include_neutrals_rbm}, using the transport settings corresponding to turquoise squares in the left panel of Figure \ref{fig:results_elmy_eda_transport}. For most discharges with $n_{e} < 2 \times 10^{20}$ m$^{-3}$, this constraint yields good prediction, as parameterizing the $n_{0}^\mathrm{sep}$ boundary condition with a wall pressure measurement better characterizes the source of fueling neutrals for these pedestals of differing density. As $n_{e} \rightarrow 2.0 \times 10^{20}$ m$^{-3}$, approximately the transition to the EDA, the model begins to overpredict the density. 

 
Two possibilities exist -- either the neutral source is overestimated or the plasma transport is underestimated. From this model alone, it is difficult to determine with certainty which is the case, but it does allow an educated guess. The former might be the case if KN1D overestimated $n_{0}^\mathrm{sep}$ for these EDAs. This could occur, for example, if $n_{e}$ was poorly characterized in the SOL such that the neutral attenuation of recycling neutrals was underestimated. At the highest densities, however, $n_{e}$ is already as high as 10$^{20}$ m$^{-3}$ in the near-SOL, and over 0.5 $\times 10^{20}$ m$^{-3}$ at the limiter, and as mentioned previously, might be overpredicted in the limiter shadow, especially at the highest densities. Increasing $n_{e}$ further leading to underestimated neutral attenuation here seems nonphysical, but without better diagnosis of the far-SOL plasma, it remains a possibility.

Alternatively,  EDA H-modes may have increased particle transport relative to ELMy H-modes that may not be captured by the standard transport model. Recent work has recognized and attempted to quantify the influence of interchange-driven RBMs on particle transport in the edge of high density discharges \cite{miller_fluxes_2025}. Following from this study, Equation \ref{eq:d_ped} is modified to include contribution from RBM-driven transport and accommodate for this additional, non-diffusive transport channel as follows:

\begin{equation}
    D_\mathrm{ped} = D_\mathrm{neo} + D_\mathrm{KBM} + D_\mathrm{TG} + D_\mathrm{RBM}
    \label{eq:d_ped_new}
\end{equation}
where $D_\mathrm{RBM}$ is the contribution from the RBM, developed from databases of previous edge particle transport studies on Alcator C-Mod \cite{miller_enhanced_2025, miller_fluxes_2025}. Two expressions for $D_\mathrm{RBM}$ are tested with this dataset. The first assumes that the RBM scales with $\alpha_{t}$, such that particle transport from the RBM is driven more strongly at high $\alpha_{t}$. This is given by:

\begin{equation}
    D_\mathrm{RBM} = C_\mathrm{RBM}\alpha_{t}^{a}
    \label{eq:d_rbm_alphat}
\end{equation}

In \cite{miller_enhanced_2025}, for a dataset at fixed $I_{P}$, it was observed that the particle transport coefficient at mid-pedestal was strongly dependent on $\alpha_{t}$ at the separatrix, with an exponent, $a = 1.2$. The dependence of $\alpha_{t}^{1.2}$ is approximately linear, which is broadly consistent with previously derived analytic expressions for RBM-driven particle transport suggesting $D_\mathrm{RBM} \sim \alpha_{d}^{-2} \sim \alpha_{t}$ \cite{mccarthy_stability_1992, manz_how_2025}, where $\alpha_{d}$ is a diamagnetic parameter introduced in previous studies of the RBM \cite{rogers_phase_1998}. The second expression follows from recent work of a larger database of C-Mod discharges, where the characteristic wavenumber for fluctuations from the RBM, $k_\mathrm{RBM}$, taken from the DALF equations, was found to better organize $\Gamma_{\perp}$ across H- and L-modes at high $n_{e}$ \cite{miller_fluxes_2025}. That study also found that additional dependence on $\hat{q}_\mathrm{cyl}$ existed by considering a dataset at variable $I_{P}$. Although it focused on $\Gamma_{\perp}$, here a form for $D_\mathrm{RBM}$ inspired by the same parameters, namely $k_\mathrm{RBM}$ and $\hat{q}_\mathrm{cyl}$ is proposed as follows:





\begin{equation}
    D_\mathrm{RBM} = C_\mathrm{RBM}^{k} \frac{1}{k_\mathrm{RBM}^{2}\hat{q}_\mathrm{cyl}}
    \label{eq:d_rbm_krbm}
\end{equation}

It should be noted that the studies from which these expressions were taken considered particle transport exclusively at mid-pedestal in the first and at the separatrix in the second. Here, the assumption is made that these expressions are valid throughout the pedestal wherever the RBM is active, which is ultimately given by the local values of $\alpha_{t}$ and $k_\mathrm{RBM}$ throughout the pedestal. To estimate the strength of transport driven by the RBM, profiles of $\alpha_{t}$ and $k_\mathrm{RBM}$ are computed across the pedestal and Equations \ref{eq:d_rbm_alphat} and \ref{eq:d_rbm_krbm} are applied across the entire profile. In reality, RBM transport may be driven more strongly near the separatrix than at the pedestal top, for example, and it may be that RBM transport onset occurs at a critical value of $\alpha_{t}$ or $k_\mathrm{RBM}$. But, where $\alpha_{t} \ll 1$ (or $k_\mathrm{RBM}$ is large), the value of $D_\mathrm{RBM}$ estimated by these expressions is small, such that inclusion of these effects may be minor.


Both of these expressions are tested in the right panel of Figure \ref{fig:results_elmy_eda_include_neutrals_rbm} through inclusion of an RBM-driven transport channel as in the previous section. Solid red squares use the same expression from Equation \ref{eq:d_rbm_alphat}, with $C_\mathrm{RBM}^{\alpha_{t}} = 0.039$, taken empirically from the dataset in \cite{miller_enhanced_2025}. Adding this transport mechanism reduces predictions as high as $n_{e}^\mathrm{ped} = 4 \times 10^{20}$ m$^{-3}$ down to more reasonable values of $n_{e}^\mathrm{ped} = 3 \times 10^{20}$ m$^{-3}$. Open red squares in the right plot use the expression from Equation \ref{eq:d_rbm_krbm}, with $C_\mathrm{RBM}^{k} = 0.086$, taken from the dataset in \cite{miller_determination_2025}. This alternate form improves the EDA prediction if slightly sacrificing accuracy in the prediction of discharges at moderate $n_{e}$. Both forms of $D_\mathrm{RBM}$, however, appear promising candidates to describe this additional resistive transport mechanism and to improve the prediction of $n_{e}^\mathrm{ped}$ for EDA H-modes at high $n_{e}$. Indeed, adding this additional transport channel appears important in extending application of the Saarelma-Connor model to high-density regimes not limited by Type-I ELMs.

\section{Comparison with predictions from EPED}
\label{sec:eped}

As mentioned in Section \ref{subsec:sc_model}, the coupled version of the Saarelma-Connor density prediction model is intended for profile prediction for Type-I ELMy H-modes, as it determines the composite $p_{e}$ profile by solving for stability to global PBMs and assuming a KBM-limited pedestal structure. To predict profiles in regimes free of Type-I ELMs, improvement of models to understand and predict turbulence-limited pedestals (rather than MHD-limited) is of great importance. The first step to doing this is to understand for which types of profiles the KBM-PBM constraint might be sufficient such that a coupled prediction might be useful and for which type an extension of models to include other constraints imposed by other instabilities might be required. In this section, scans with the EPED code \cite{snyder_development_2009} are performed, using new simulation capabilities for analysis of C-Mod and prediction of SPARC pedestals \cite{han_pedestal_2025}, to assess the validity of the KBM-PBM constraint in determining the pedestal structure of the experimental dataset, including H-modes with and without Type-I ELMs. Details of the setup of these simulations can be found in \ref{sec:eped_details}.

\begin{figure}
\centering
\includegraphics[width=0.9\columnwidth]{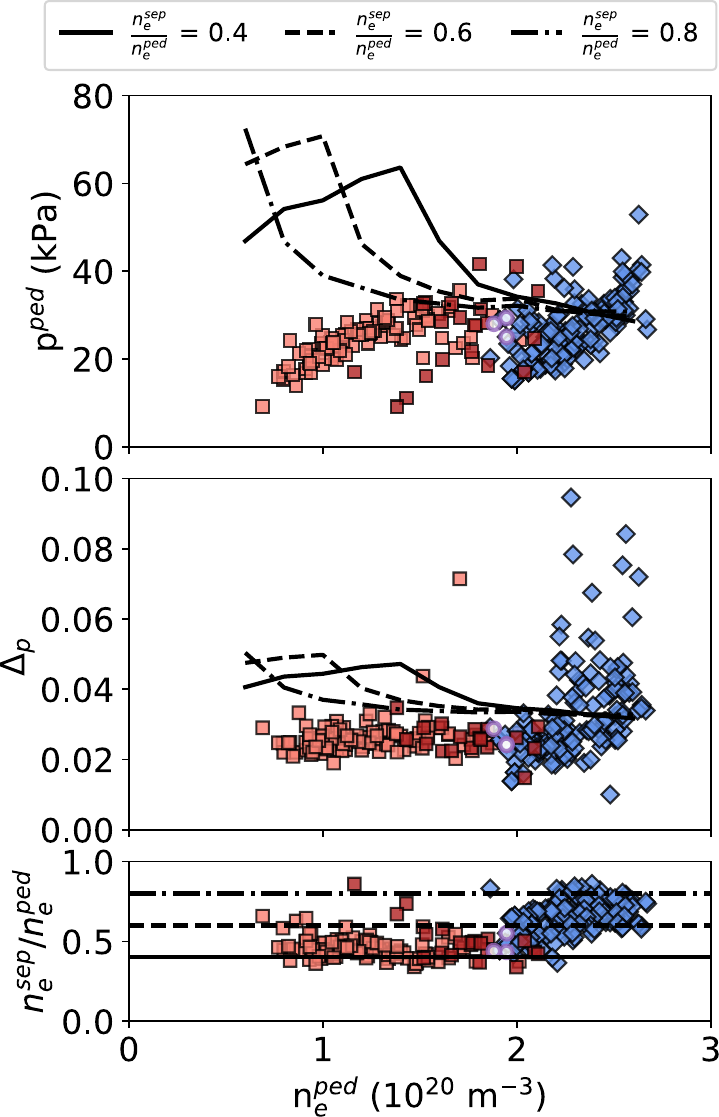}
\caption{Total pedestal top pressure (top), pressure pedestal width (middle), and ratio of separatrix to pedestal density (bottom) as a function of pedestal density. Individual markers in all three plots correspond to experimental data and use the categorization in previous sections: discharges with small ELMs (light red squares), large ELMs (dark red squares), mixed EDA and ELMy (purple circles), and only EDA (blue diamonds). In the top two plots, EPED results are shown using lines for three values of $\frac{n_{\mathrm{sep}}}{n_{\mathrm{ped}}}$ = 0.4, 0.6, and 0.8, in solid, dashed, and dash-dotted curves, respectively. At bottom, horizontal lines represent the three values in $\frac{n_{\mathrm{sep}}}{n_{\mathrm{ped}}}$ corresponding to the EPED scans, using the same line styles as above.} 
\label{fig:eped_results}
\end{figure}

The results of this modeling exercise are shown in Figure \ref{fig:eped_results}. The top two panels show the results of the EPED prediction for the height (here plotted as $p_{e}^\mathrm{ped}$, rather than $\beta_{p}^\mathrm{ped}$) and $\Delta_{p}$ of the marginally KBM-PBM stable profiles, as a function of the input $n_{e}^\mathrm{ped}$. It shows the results for the three scans in $n_{e}^\mathrm{ped}$, each with a different value of $\frac{n_{e}^\mathrm{sep}}{n_{e}^\mathrm{ped}}$, at 0.4, 0.6, and 0.8, in solid, dashed, and dash-dotted lines respectively. Each set of EPED solutions at a given $\frac{n_{e}^\mathrm{sep}}{n_{e}^\mathrm{ped}}$ has a branch at low $n_{e}$, the peeling-limited branch, where the height (and the width) increase with $n_{e}^\mathrm{ped}$ and a branch at high $n_{e}$, where these quantities decrease with $n_{e}^\mathrm{ped}$. The transition between these two branches is where the maximum predicted $p^\mathrm{ped}$ is obtained, occurring at the peeling-ballooning (PB) transition density, $n_{e}^\mathrm{ped, PB}$. Note that for $\frac{n_{e}^\mathrm{sep}}{n_{e}^\mathrm{ped}} = 0.8$, the results all lie on the ballooning branch for the simulated density range, although peeling-limited solutions presumably exist for $n_{e}^\mathrm{ped} < 0.6 \times 10^{20}$ m$^{-3}$.



\subsection{Comparison with experiment}
\label{subsec:results_comparison}
The top two panels of Figure \ref{fig:eped_results} also show the experimentally-determined values of $p^\mathrm{ped} = 2p_{e}^\mathrm{ped}$, using the assumption that $T_{e} = T_{i}$, as well as the experimental $\Delta_{p} = \Delta_{p_{e}}$ determined from a fit to the $p_{e}$ profile directly \footnote{Note that some works use the approximation $\Delta_{p} = (\Delta_{n} + \Delta_{T})/2$, with $\Delta_{n}$ and $\Delta_{T}$ the widths of the fits to the $n_{e}$ and $T_{e}$ profiles, respectively. The current approach to directly fit the $p_{e}$ profile is found to yield less scatter but also a slightly narrower width, which should be kept in mind when comparing to previously-reported values of $C$, for example. The ratio between the widths using the different approaches, however, is relatively constant across the dataset, such that the width-height dependence is unchanged regardless of choice of width determination.}. The bottom panel of Figure \ref{fig:eped_results} shows the ratio $\frac{n_{e}^\mathrm{sep}}{n_{e}^\mathrm{ped}}$ determined experimentally, also as a function of $n_{e}^\mathrm{ped}$. Horizontal lines show values of fixed $\frac{n_{e}^\mathrm{sep}}{n_{e}^\mathrm{ped}} = 0.4, 0.6, 0.8$ as a reference to allow easier interpretation of the above plot with the EPED results. Rather than a direct validation of the code or interpretation of one particular experiment, both of which have been published using EPED on C-Mod \cite{walk_characterization_2012, hughes_pedestal_2013, hughes_access_2018, snyder_high_2019}, this exercise is meant to provide high-level insight for how these different H-modes compare with expectations from MHD instability theory for Type-I ELMs. Without even considering the experimental data, the first thing that is clear from scanning $\frac{n_{e}^\mathrm{sep}}{n_{e}^\mathrm{ped}}$ is that indeed, this ratio matters for the prediction, especially at lower and middle densities. Experimentally, the ratio is anywhere between 0.4 -- 0.5 for ELMy H-modes but increases to 0.8 for EDA H-modes. 

Looking again only at the EPED solutions, the effect of increasing $\frac{n_{e}^\mathrm{sep}}{n_{e}^\mathrm{ped}}$ is dual. The first effect is to move $n_{e}^\mathrm{ped, PB}$ to lower densities, such that as noted earlier, for $\frac{n_{e}^\mathrm{sep}}{n_{e}^\mathrm{ped}} = 0.8$, there are no peeling-limited solutions found for the range of $n_{e}^\mathrm{ped}$ scanned. The second effect is that the maximum attainable pressure, $p^\mathrm{ped,max} = p^\mathrm{ped}|_{n_{e}^\mathrm{ped, PB}}$, increases with $\frac{n_{e}^\mathrm{sep}}{n_{e}^\mathrm{ped}}$. The combination of these effects yields the finding that if a pedestal is ballooning-limited, increasing $\frac{n_{e}^\mathrm{sep}}{n_{e}^\mathrm{ped}}$ decreases $p_{e}^\mathrm{ped}$, while the opposite is true for a peeling-limited pedestal. This second effect, however, is only important for $n_{e}^\mathrm{ped} < 10^{20}$ m$^{-3}$, whereby a higher pressure might exist at higher $\frac{n_{e}^\mathrm{sep}}{n_{e}^\mathrm{ped}}$ if the pedestal is still peeling-limited, assuming of course erroneously that $n_{e}^\mathrm{ped}$ and $n_{e}^\mathrm{sep}$ can be modified independently experimentally. Few pedestals on C-Mod exist below $10^{20}$ m$^{-3}$ such that for the range of relevant $n_{e}^\mathrm{ped}$, increasing $\frac{n_{e}^\mathrm{sep}}{n_{e}^\mathrm{ped}}$ is generally detrimental to pedestal performance within the framework of ideal MHD. Similar results from stability calculations have been obtained for a change to $\frac{n_{e}^\mathrm{sep}}{n_{e}^\mathrm{ped}}$ triggered by a relative outward shift in the $n_{e}$ profile relative to the $T_{e}$ profile instead of a flattening of the $n_{e}$ profile \cite{dunne_role_2017, frassinetti_role_2019}. At the highest values of $n_{e}^\mathrm{ped}$, which are indeed more typical for C-Mod, the EPED-predicted solutions converge, and $p^\mathrm{ped} \rightarrow 30$ kPa, becoming independent of $\frac{n_{e}^\mathrm{sep}}{n_{e}^\mathrm{ped}}$. Note that all of the above conclusions also apply to the width prediction shown in the center panel of Figure \ref{fig:eped_results}, as the width-height scaling is positive and monotonic.



Sensitivity to $\frac{n_{e}^\mathrm{sep}}{n_{e}^\mathrm{ped}}$ means comparison with experiment requires nuance. Beginning first at the lowest $n_{e}^\mathrm{ped}$ with the small ELM discharges in light red, $\frac{n_{e}^\mathrm{sep}}{n_{e}^\mathrm{ped}} = 0.4 - 0.5$. Here, one needs to consider the solid curves, or perhaps interpolate onto one somewhere between solid and dashed. Experimentally, in this range of $n_{e}^\mathrm{ped}$, $p^\mathrm{ped}$ increases with $n_{e}^\mathrm{ped}$, a hallmark of peeling-limited pedestals. Regardless of which set of EPED solutions one picks to compare against (solid or dashed), it is clear that the prediction for $p^\mathrm{ped}$ from EPED is considerably higher than the experimental value. These pedestals are likely not at the PB boundary. Some other mechanism then may be responsible for limiting the height of these pedestals. With their prominent $T_{e}$ pedestals, these profiles resemble I-modes (Figure \ref{fig:elmy_eda_profiles}), which have also been shown to live far from the PB boundary \cite{hughes_pedestal_2013}. Regardless, they also have an $n_{e}$ pedestal, exist in the favorable drift direction, and do not appear to feature the WCM often associated with the I-mode, all characteristics that catalog them as H-modes.


As $n_{e}^\mathrm{ped} \rightarrow 1.4 \times 10^{20}$ m$^{-3}$, the ELMs grow larger and discharges move into the ``large ELMs'' categorization, shown as dark red squares. Experimentally, $p^\mathrm{ped}$ (and $\Delta_{p}$) ceases to increase with $n_{e}^\mathrm{ped}$ and even begins to decrease with $n_{e}^\mathrm{ped}$. This behavior is often linked with the transition from the peeling to the ballooning branch. Indeed, for the solutions at $\frac{n_{e}^\mathrm{sep}}{n_{e}^\mathrm{ped}} = 0.4$, this occurs at $n_{e}^\mathrm{ped,PB} \approx 1.4 \times 10^{20}$ m$^{-3}$, with the experimental maximum value only slightly higher at $\sim$1.6$\times 10^{20}$ m$^{-3}$. Large ELM data suffer somewhat more scatter than small ELM and EDA data, so it is difficult to compare them to the simulations. Between $n_{e}^\mathrm{ped} = 1.5 - 1.9\times 10^{20}$ m$^{-3}$, however, EPED finds these pedestals on the ballooning branch, in line with previous EPED and more detailed ELITE modeling of Type-I ELMy H-modes on C-Mod \cite{hughes_pedestal_2013}. In this range, the predicted height matches that of some large ELM experimental points, with the curve at $\frac{n_{e}^\mathrm{sep}}{n_{e}^\mathrm{ped}} = 0.6$ offering a slightly improved match. Note that EPED-predicted widths converge to $\sim$0.035 and are generally larger than the experimentally-computed widths, which are typically below 0.03, even for large ELM discharges with good height predictions. Taken together, these findings are generally consistent with previous EPED interpretive modeling on C-Mod done for plasmas in very similar shape and with very similar equilibrium parameters \cite{hughes_access_2018}.


\subsection{Deviations from EPED in the EDA H-mode}
\label{subsec:eda_stability}

The last and possibly most intriguing observation from this exercise is the comparison with EDAs. As $n_{e}^\mathrm{ped}$ continues to grow, discharges transition to EDA and the ratio $\frac{n_{e}^\mathrm{sep}}{n_{e}^\mathrm{ped}} \rightarrow 0.8$, as seen in Section \ref{sec:exp_profs_neutrals}. The densities at which EDAs appear are all solidly in the ballooning branch and regardless of which black curve one picks, the EPED prediction would give $p^\mathrm{ped} \approx 35-40$ kPa were the discharge a Type-I ELMy H-mode. Experiment, however, would suggest that it may be difficult to find a Type-I ELMy H-mode with large enough pressure gradients to still be KBM/PBM-limited at high $\frac{n_{e}^\mathrm{sep}}{n_{e}^\mathrm{ped}}$. Indeed, these EDA H-modes are not expected to be limited by this physics and agreement with EPED is not expected -- comparison is only meant to be instructive, not a comment on model validity. For many H-modes in the dataset, the EPED prediction, even at the highest values of $n_{e}^\mathrm{ped}$ and $\frac{n_{e}^\mathrm{sep}}{n_{e}^\mathrm{ped}}$, is larger than the experimental height and width, with many EDA H-modes at or below 30 kPa. There are, however, a non-negligible number of EDA H-modes that have a pressure \emph{larger} than the EPED prediction. 

\begin{figure}
\centering
\includegraphics[width=\columnwidth]{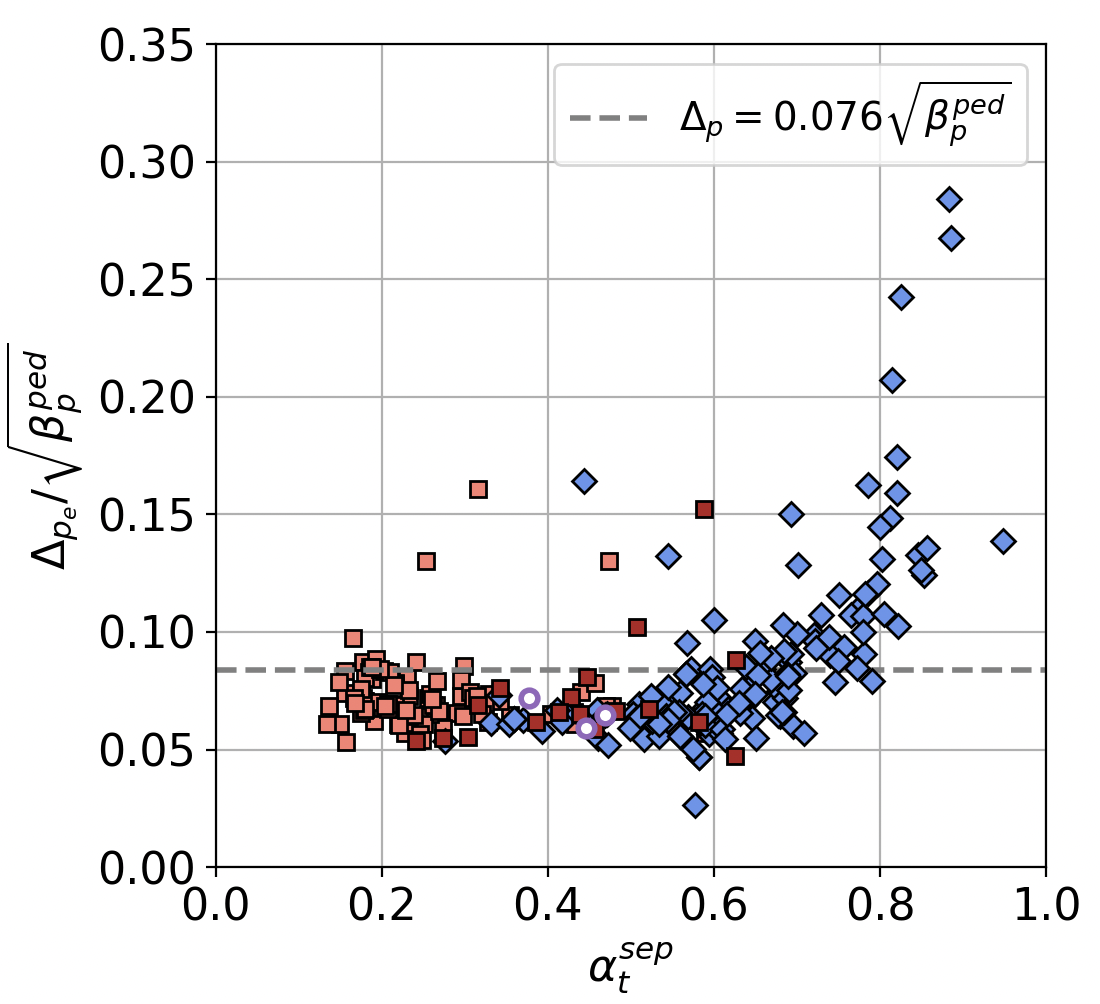}
\caption{Experimental ratio of $\Delta_{p}$ to $\sqrt{\beta_{p}^\mathrm{ped}}$ against $\alpha_{t}^\mathrm{sep}$ using the categorization in previous sections: discharges with small ELMs (light red squares), large ELMs (dark red squares), mixed EDA and ELMy (purple circles), and only EDA (blue diamonds). The dashed curve in gray corresponds to a proportionality constant, $C = 0.076$.} 
\label{fig:kbm_broadening}
\end{figure}

This may at first seem surprising considering previous work on C-Mod that showed an EDA H-modes far from the PB boundary as calculated by ELITE \cite{hughes_pedestal_2013}, whereas a Type-I ELMy H-modes lived close to the boundary, as predicted by PB theory. A closer look at Figure 6 of \cite{hughes_pedestal_2013}, for example, shows that the EDA H-mode phase considered actually has \emph{higher} normalized pressure gradient. Its larger distance to the ballooning boundary relative to the ELMy H-mode, results not from the EDA moving to lower $\alpha$, but rather the boundary moving to higher $\alpha$ for the EDA. Possibly, this is an indication of the stronger sensitivity of Type-I ELMy than EDA H-modes to shaping. In these more weakly shaped plasmas, it may be that the pedestal performance of ELMy H-modes is significantly degraded relative to that of the EDA H-modes, which may be more robust to changes to shaping. Whatever the explanation, these data suggest that it is not obvious that in all scenarios, all EDA H-modes perform more poorly than ELMy H-modes. 

To try to understand this unexpected result, Figure \ref{fig:kbm_broadening} probes the KBM width-height EPED assumption for the experimental pedestal profiles. It plots the ratio of the pedestal width to the height against $\alpha_{t}^\mathrm{sep}$, the parameter used throughout this work parameterizing the drive of resistively-driven transport. Plotted also is a horizontal line depicting the typically-assumed proportionality coefficient, $C = 0.076$. For most discharges in the dataset, including some EDA H-modes at moderate values of $\alpha_{t}$, $\Delta_{p}/\sqrt{\beta_{p}^\mathrm{ped}}$ is fairly constant, and only slightly below $0.076$. Their pedestal structure matches that of a KBM-limited pedestal, although of course some are also PBM-limited and feature ELMs, while others do not. For ELMy discharges, this finding is consistent with previous fluctuation analysis of this dataset suggesting that inter-ELM transport of ELMy discharges is indeed limited by an instability with the KBM signature \cite{diallo_observation_2014, diallo_correlations_2015}. At the highest values of $\alpha_{t}$, however, there is a clear deviation from the scaling. These high $n_{e}$ pedestals are wider given their height than the KBM scaling would predict, implying that there is some other instability that becomes destabilized at high $\alpha_{t}$ which acts to broaden the pedestal. This is strongly reminiscent of the broadening of local separatrix gradient scale lengths with $\alpha_{t}$ observed on multiple devices, including AUG \cite{eich_turbulence_2020}, C-Mod \cite{labombard_evidence_2005, miller_determination_2025}, and EAST \cite{li_study_2025}. 

As identified in those works, a candidate for this broadening is a resistively-driven mode, like the RBM. A number of possibilities exist here. It is possible, that the RBM (or something scaling like it) altogether supplants the KBM, instead of simply driving transport additionally to it, as the proposed form of Equation \ref{eq:d_ped_new} would suggest. It is also possible that at low values of $\alpha_{t}$, the KBM triggers the DAW as found in \cite{grenfell_multi-faced_2024}, but at high values of $\alpha_{t}$, the KBM instead triggers RBM turbulence, which may also spread to the pedestal, as suggested in \cite{manz_how_2025}. Or it may be that RBMs are never excited and that the changes to pedestal structure can be understood entirely through the inclusion of non-ideal MHD effects, which become more important at high resistivity, as was recently shown for JET pedestals \cite{stefanelli_non-ideal_2025}. More detailed simulations with a visco-resistive stability code like CASTOR3D \cite{puchmayr_nodate} or fluid turbulence simulations with a code like GRILLIX \cite{zholobenko_electric_2021}, would be of great interest to confirm the role of resistivity and probe the physics of these non-ELMing H-modes.


The experimental analysis and modeling shown in this and the previous section provide a compelling argument that understanding pedestals, specifically transport-limited, high-density pedestals, requires consideration of the entire pedestal profile, including crucially its value at the separatrix. There is recent evidence that in the case of the QCE, driving the separatrix locally ballooning-unstable limits the pressure gradient such that it does not reach this limit \cite{radovanovic_developing_2022, dunne_quasi-continuous_2024}. Similar profile-limiting modes must be present also for EDAs \cite{cathey_probing_2023}, which feature similar profile characteristics to the QCE, although possibly at a different location in the profile. To complicate matters further, not all EDA H-modes on C-Mod appear to be created equal. Some exist at pressures higher than the EPED-predicted height and others below. Some have a structure that is consistent with a KBM-limited profile -- others clearly deviate and widen at high values of $\alpha_{t}$. The modeling performed in the previous two sections and the experimental analysis from the first two sections suggests that especially in the case of the EDA H-mode, understanding the interaction between stability to global and local modes with the mechanisms that determine the density pedestal, be they fueling or transport, is crucial to understanding the pedestal. Ultimately, a successful pedestal model must couple local conditions, like those at the separatrix, with the global pedestal state to understand the interplay between turbulent transport and MHD stability across different H-mode regimes.

\section{Initial predictions of the SPARC pedestal density}
\label{sec:sparc_prediction}

\begin{figure}[t]
\centering
\includegraphics[width=0.9\columnwidth]{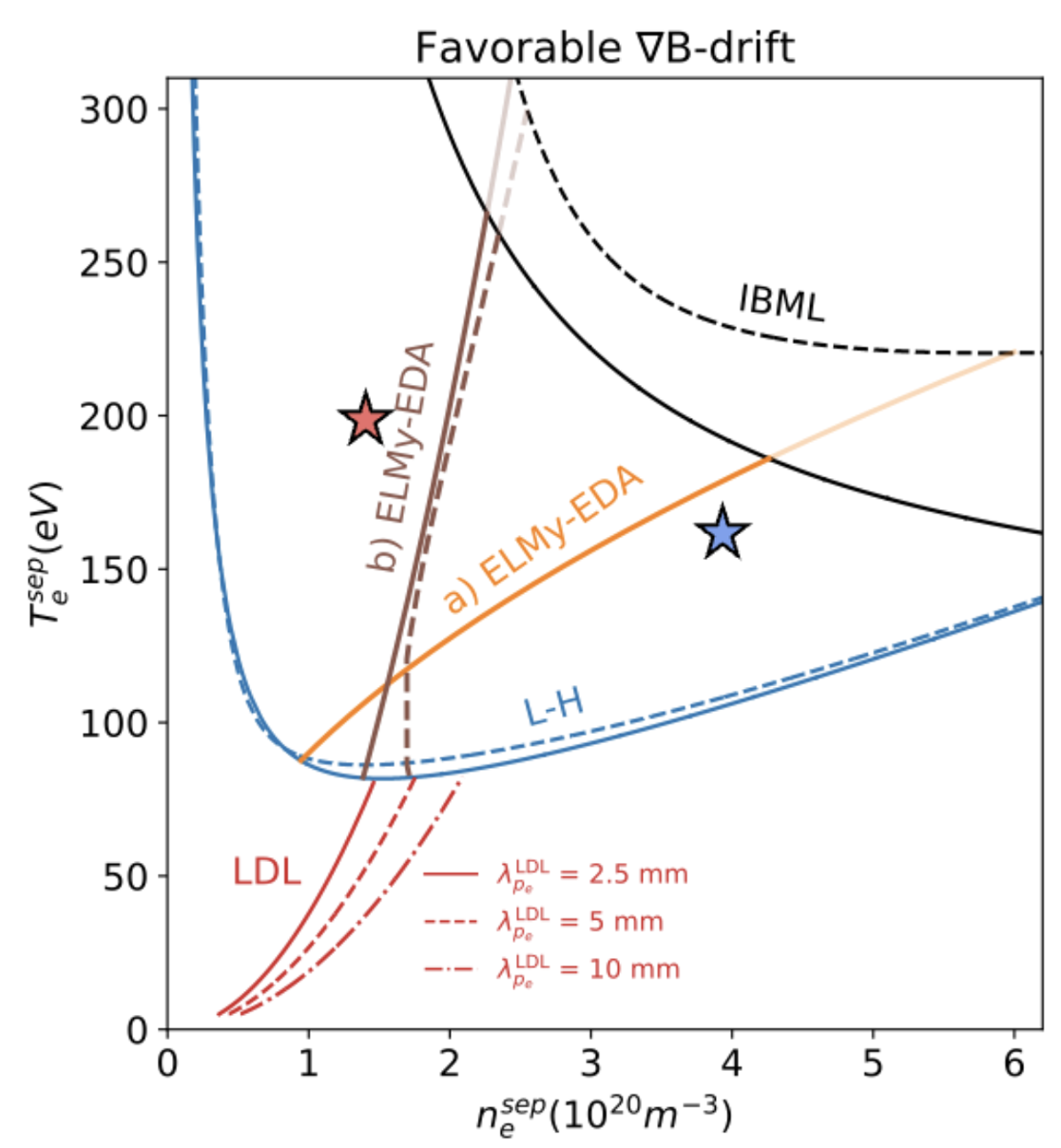}
\caption{Projected boundaries for the separatrix operational space of SPARC based on PRD parameters using C-Mod (solid) and AUG (dashed) $\lambda_{p_{e}}$ scalings. Primary SepOS boundaries as well as proposed ELMy-EDA boundaries are shown. Red star shows operational point for PRD. Blue star shows high density operational point.}
\label{fig:sparc_sepos}
\end{figure}

The final section of this paper leverages the modeling and observations from earlier sections to make predictions for the SPARC edge plasma. The SepOS prediction of the operational space of SPARC is first used to generate a boundary condition in $n_{e}^\mathrm{sep}$, as well as to estimate important profile quantities like $\alpha_{t}$, $\beta_{e}$, and $\lambda_{p_{e}}$. This was done in \cite{miller_determination_2025} for both the favorable and unfavorable drift directions, but for the purposes of model application, only the favorable drift direction dataset will be considered. Figure \ref{fig:sparc_sepos} shows the separatrix operational space in the favorable drift direction for SPARC. Its PRD scenario \cite{creely_overview_2020, Rodriguez-Fernandez_2022, body_sparc_nodate}, designed to reach high fusion gain, $Q\gg 1$, is designed using the set of engineering parameters listed in Table \ref{tab:sparc_params}. From these engineering parameters, DALF-normalized quantities are calculated and combinations of these quantities are numerically solved for SPARC's parameters to generate the SepOS boundaries. These are determined for the L-mode density limit (LDL), the L-H transition (L-H), and the ideal MHD ballooning limit (IBML) in red, blue, and black respectively. Additionally, two other boundaries are shown from previous C-Mod and AUG work \cite{miller_determination_2025, faitsch_analysis_2023} for transition criteria at the separatrix between ELMy H-modes and a high density H-mode regime free of Type-I ELMs (the EDA on C-Mod and the QCE on AUG). Stars represent two operational points proposed for SPARC operation, which will be detailed  below.

\begin{table}[h]
\begin{center}
\caption{Parameters for SPARC's primary reference discharge \cite{creely_overview_2020, Rodriguez-Fernandez_2022, body_sparc_nodate}}
\label{tab:sparc_params}
\begin{tabular}{cc}
Parameter & Value \\
\midrule
\midrule
$B_{t}$ (T) & 12.2 \\
\midrule
$I_{P}$ (MA) & 8.7 \\
\midrule
$B_{p}$ (T) & 1.7 \\
\midrule
R (m) & 1.85 \\
\midrule
a (m) & 0.57 \\
\midrule
$\delta$ & 0.54 \\
\midrule
$\kappa$ & 1.97 \\
\midrule
$\langle n_{e} \rangle$ (10$^{20}$m$^{-3}$) & 2.9 \\
\midrule
$f_{G}$ & 0.34 \\
\midrule
\end{tabular}
\end{center}
\end{table}

\begin{figure*}
\centering
\includegraphics[width=1.6\columnwidth]{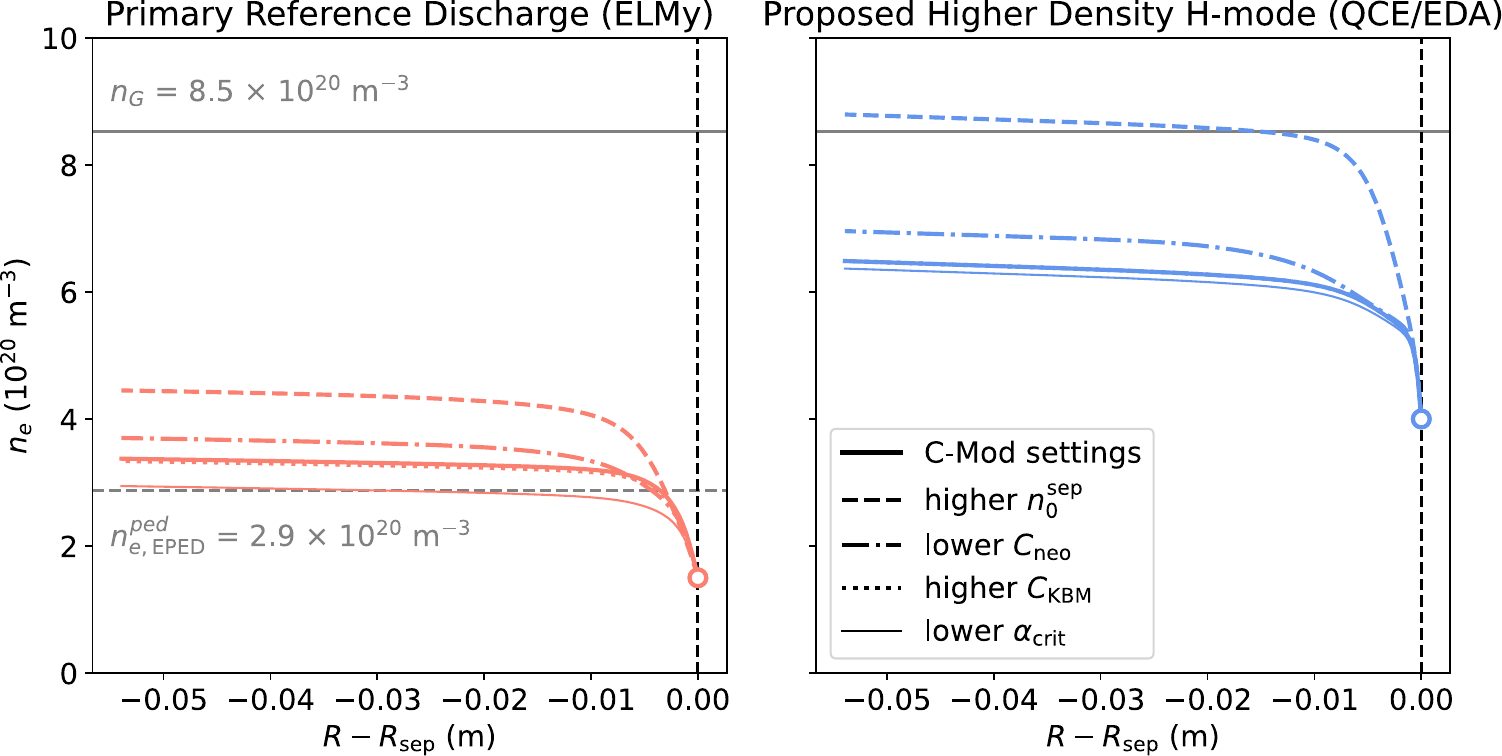}
\caption{Prediction of SPARC pedestal density profile for PRD scenario (left) and high-density H-mode (right), using standard settings from C-Mod validation (thick, solid), higher $n_{0}^\mathrm{sep} = 10^{16}$ m$^{-3}$ (dashed), lower $C_\mathrm{neo} = 0.005$ (dash-dotted), higher $C_\mathrm{KBM} = 0.1$ (dotted), and lower $\alpha_\mathrm{crit} = 2$ (thin, solid). Shown also are the Greenwald density, $n_{G}$ and the input to the EPED prediction, $n_{e,\mathrm{EPED}}^\mathrm{ped}$.}
\label{fig:sparc_density_prediction}
\end{figure*}

With boundaries for allowable separatrix parameters on SPARC, it becomes possible to similarly project achievable pedestal parameters using a prediction model like the Saarelma-Connor model. This is done for two previously-proposed operational scenarios for SPARC, including also the sensitivity to some of the model's free parameters. The two operational points projected are first, the PRD, mentioned already in this section, and second, a recently-proposed high density operation point \cite{eich_separatrix_2024}. The PRD has been extensively studied and modeled \cite{hughes_projections_2020, rodriguez-fernandez_predictions_2020, rodriguez-fernandez_nonlinear_2022, Rodriguez-Fernandez_2022} in previous work. Predicted pedestal $T_{e}$ (as well as core $n_{e}$ and $T_{e}$) profiles have been taken from these. The results are of course not expected to be self-consistent, since there is no iteration back to these original predictions. The higher density operational point has only recently been proposed, with view of ameliorating SPARC's power handling challenge, and core and pedestal stability and transport predictive capabilities have not as of yet been leveraged for this scenario. Its input $T_{e}$ profile will be based off the PRD prediction, modified using knowledge of the EDA H-mode on C-Mod, as will be explained below.


\begin{figure*}
\centering
\includegraphics[width=1.6\columnwidth]{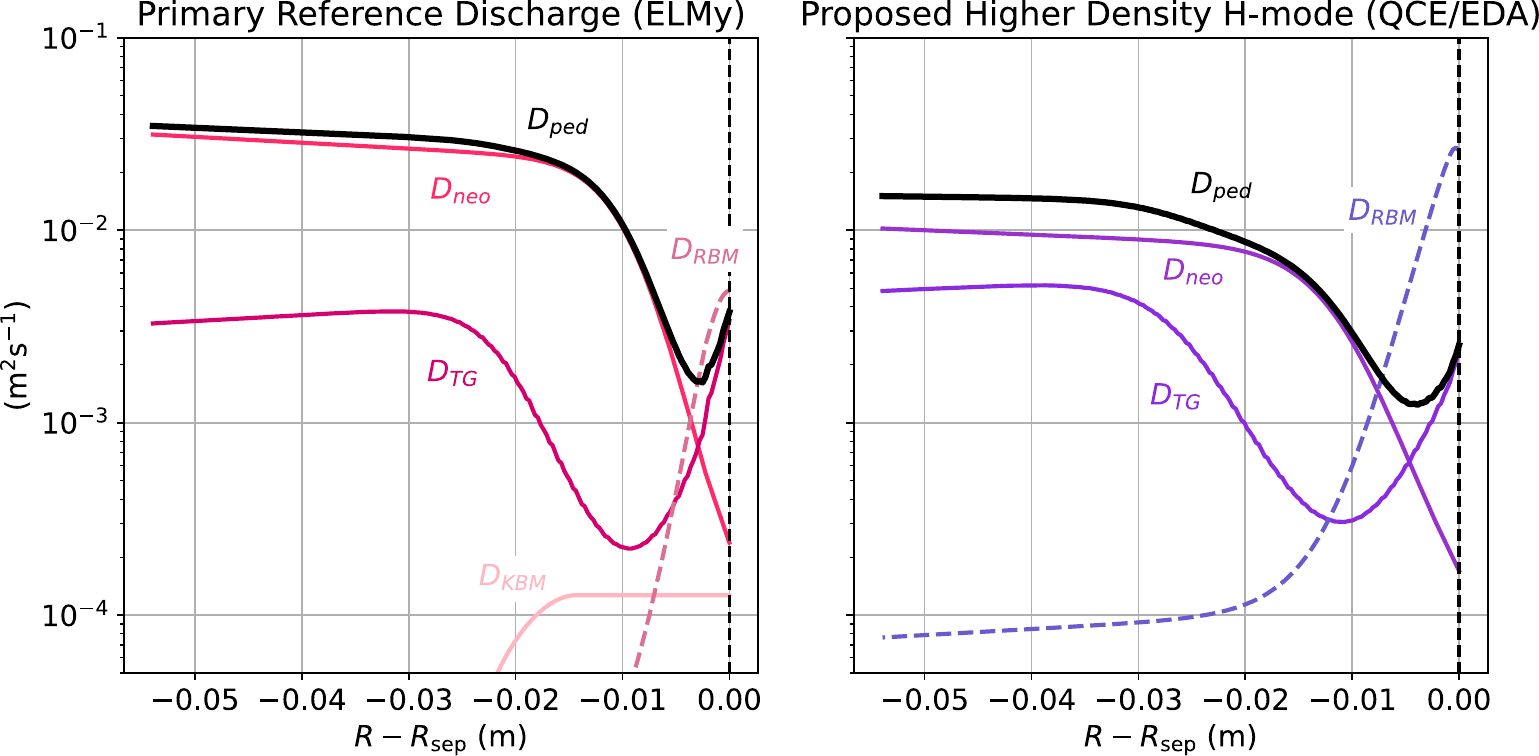}
\caption{Modeled $D$ profile of density pedestal prediction of SPARC for PRD scenario (left) and high-density H-mode (right), showing total $D_\mathrm{ped}$ in black, as well as $D_\mathrm{neo}$, $D_\mathrm{KBM}$, $D_\mathrm{TG}$, and $D_\mathrm{RBM}$ (not included in the total $D_\mathrm{ped}$ for initial simulations) in shades of pink and purple.}
\label{fig:sparc_density_prediction_transport}
\end{figure*}

Figure \ref{fig:sparc_sepos} shows the two proposed operating points with stars -- the red star corresponds to the low $n_{e}$ PRD and the blue star corresponds to the high $n_{e}$ point. The PRD is asserted to operate at $n_{e}^\mathrm{sep} = 1.5 \times 10^{20}$ m$^{-3}$ and $T_{e}^\mathrm{sep} = 195$ eV. This puts this scenario in the Type-I ELMy H-mode space. The high density operation point has been proposed at $n_{e}^\mathrm{sep} = 4.0 \times 10^{20}$ m$^{-3}$ and $T_{e}^\mathrm{sep} = 156$ eV. This point still lies in the H-mode operational space, but to the right of either proposed boundary for disappearance of Type-I ELMs, placing it in the EDA/QCE space. The expectation is then that this scenario would not feature large ELMs and would as such, not be limited by the PBM-KBM constraint from EPED, making prediction of pedestal parameters difficult. Without previously computed predictions for pedestal widths and heights, or without core transport profiles generated through gyrokinetic simulation, very crude approximations are used to supply the Saarelma-Connor model with $T_{e}$ and $\alpha$ profiles with which to compute $D_\mathrm{ped}$. Assuming these points are in ELMy and EDA regimes respectively, trends from the ELMy-EDA dataset from C-Mod are used to extrapolate approximate changes to pedestal characteristics and core profiles. Very simply the core $T_{e}$ profile is divided by 2, i.e. $T_{e}^\mathrm{EDA/QCE} = \frac{1}{2}T_{e}^\mathrm{PRD}$ and the core $n_{e}$ profile is multiplied by 1.5, i.e. $n_{e}^\mathrm{EDA/QCE} = 1.5n_{e}^\mathrm{PRD}$. These choices are motivated by comparison of pedestal top values from the bottom plots of Figure \ref{fig:dataset_moving_averages}. As far as pedestal profiles, EPED calculations for the PRD \cite{hughes_projections_2020} gave $p^\mathrm{ped} = 324$ kPa, $n_{e}^\mathrm{ped} = 2.9 \times 10^{20}$ m$^{-3}$, and $T_{e}^\mathrm{ped} = 3.9$ keV, with $\Delta_{p} = 0.034$ in $\psi_{n}$. For the high density discharge, pedestal profiles are generated using a slightly larger $\Delta_{p}^\mathrm{EDA/QCE} = 1.5 \Delta_{p}^\mathrm{PRD} = 0.042$, a slightly wider temperature pedestal, $\Delta_{T}^\mathrm{EDA/QCE} = 1.3 \Delta_{T}^\mathrm{PRD}$, and a slightly smaller density pedestal width, $\Delta_{n}^\mathrm{EDA/QCE} = \Delta_{n}^\mathrm{PRD}/1.3$. These rough estimates for how the pedestal widths change in the EDA compared to the ELMy H-mode are taken from the ratio of the median values of $\Delta^\mathrm{EDA}$ to those of $\Delta^\mathrm{ELMy}$ for both $n_{e}$ and $T_{e}$ for the C-Mod ELMy-EDA dataset presented above. As Figure \ref{fig:kbm_broadening} has motivated, however, the pedestal structure can vary considerably in the EDA and these approximations are only meant to be a starting point for the modeling effort.

\begin{figure*}
\centering
\includegraphics[width=1.6\columnwidth]{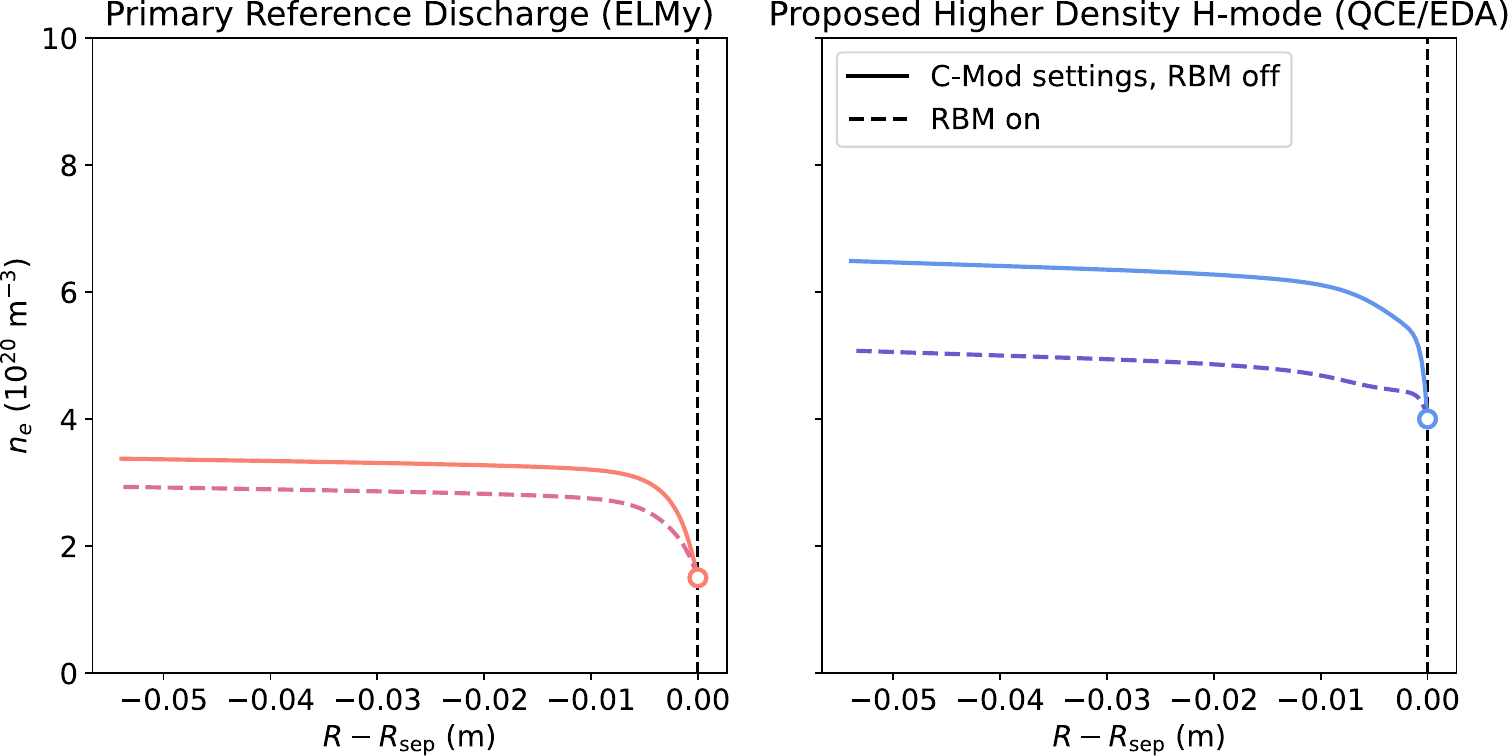}
\caption{Comparison of prediction of SPARC density pedestal profile for PRD scenario (left) and high-density H-mode (right) with RBM off (solid, red and blue) and on (dashed, pink and purple).}
\label{fig:sparc_rbm}
\end{figure*} 

Figure \ref{fig:sparc_density_prediction} shows initial results for these two scenarios, including also variation in choices for the transport and neutrals model. The density prediction model is run first with the same transport choices from the best fit to the ELMy-EDA dataset, i.e. $\left( \frac{D}{\chi} \right)_\mathrm{TG} = 0.05$, $C_\mathrm{KBM} = 0.01$, $\alpha_\mathrm{crit} = 3$, and $n_{0}^\mathrm{sep} = 10^{15}$ m$^{-3}$, yielding the thick, solid curves. For the PRD, the density prediction model predicts $n_{e}^\mathrm{ped} = 3.4 \times 10^{20}$ m$^{-3}$, somewhat above, but close to the input value of $n_{e}^\mathrm{ped}$ to EPED that yielded the predicted $p^\mathrm{ped}$, shown as a dashed gray line in the left panel of Figure \ref{fig:sparc_density_prediction}. Note, however, that those predictions were made with a lower $n_{e}^\mathrm{sep}$, so a self-consistent solution should not be expected, especially since the $\frac{n_{e}^\mathrm{sep}}{n_{e}^\mathrm{ped}}$ ratio used in those simulations differed from that assumed in these.

The figure also shows sensitivity of the prediction to model inputs. The first of these, the dashed curve, uses a higher $n_{0}^\mathrm{sep} = 10^{16}$ m$^{-3}$. Simulations with UEDGE showed that $n_{0}^\mathrm{sep}$ may even be as high as $2 \times 10^{17}$ m$^{-3}$ at the OMP \cite{ballinger_simulation_2021}, in line with values measured on Alcator C-Mod \cite{miller_enhanced_2025, miller_particle_2025}. As with C-Mod ELMy discharges, the prediction is quite sensitive to this choice, with $n_{e}^\mathrm{ped} = 4.4 \times 10^{20}$ m$^{-3}$ predicted if using the higher $n_{0}^\mathrm{sep}$. To motivate the sensitivity to parameters constraining the plasma transport model, Figure \ref{fig:sparc_density_prediction_transport} shows the overall plasma transport coefficient, $D_\mathrm{ped}$, profile in black, as well as its contributions in shades of pink. Note that although $D_\mathrm{RBM}$ is shown in the figure, it is not used in the prediction in Figure \ref{fig:sparc_density_prediction} (and is not included in the black curve shown). From Figure \ref{fig:sparc_density_prediction_transport}, $D_\mathrm{ped}$ is high throughout the pedestal and only begins to decrease a couple of cm inside the separatrix. Furthermore,  when inspecting the contributions to $D_\mathrm{ped}$, $D_\mathrm{neo}$ composes the majority of the transport profile, at least inside the mid-pedestal. Considering Equation \ref{eq:d_neo}, $D_\mathrm{neo}$ scales with $\frac{\rho_{s}^{2}c_{s}}{a}$, with $C_\mathrm{neo} = 0.05$ a constant. This expression is taken from transport simulations of DIII-D pedestals \cite{guttenfelder_testing_2021}. Given much higher $T_{e}$ on SPARC, $\rho_{s}$ and especially $c_{s}$ will be considerably higher. Since SPARC is a rather compact device, the ratio $\frac{\rho_{s}^{2}c_{s}}{a}$ is also much higher than on other devices, yielding a fairly large predicted $D_\mathrm{neo}$. Note also that $D_\mathrm{KBM}$ is exceedingly small, likely an underestimate of transport in these profiles, especially given that the PRD is expected to be unstable to Type-I ELMs. Here, considering Equation \ref{eq:d_kbm}, modifying either $C_\mathrm{KBM}$ or $\alpha_\mathrm{crit}$ would change the predicted $D_\mathrm{KBM}$. Sensitivity to these assumptions of neoclassical and KBM transport is shown also in Figure \ref{fig:sparc_density_prediction}. The prediction surprisingly increases only slightly, to $n_{e}^\mathrm{ped} = 3.7 \times 10^{20}$ m$^{-3}$, even when neoclassical transport is decreased by a factor of 10. It appears that modifying transport further inside the pedestal, which is where $D_\mathrm{neo}$ dominates, has a relatively small effect on the gradients. Increasing $C_\mathrm{KBM}$ by a factor of 10, shown in the dotted pink line has almost no effect on the prediction. This is not entirely surprising, given that $D_\mathrm{KBM}$ is already as low as $10^{-4}$ m$^{2}$s$^{-1}$, such that even a factor of 10 increase has almost no effect on the prediction. Interestingly, decreasing $\alpha_\mathrm{crit}$ from 3 to 2 has a larger effect, lowering the predicted $n_{e}^\mathrm{ped}$ to $2.9 \times 10^{20}$ m$^{-3}$, even for the low value of $C_\mathrm{KBM} = 0.01$. These sensitivities indicate that identifying the critical $\alpha$ of KBM onset may be just as important as determining the strength of the transport it drives.

The prediction looks somewhat different for the high density discharges, shown on the right of Figures \ref{fig:sparc_density_prediction} and \ref{fig:sparc_density_prediction_transport}. For standard ``C-Mod'' plasma transport settings but $n_{0}^\mathrm{sep} = 10^{15}$ m$^{-3}$, $n_{e}^\mathrm{ped} = 6.5 \times 10^{20}$ m$^{-3}$. Interestingly, when increasing $n_{0}^\mathrm{sep}$ to $10^{16}$ m$^{-3}$, the prediction increases substantially to  $8.8 \times 10^{20}$ m$^{-3}$, above the Greenwald density, $n_{G} = 8.5 \times 10^{20}$ m$^{-3}$. There is some sensitivity of the prediction to $C_\mathrm{neo}$, but almost none to the parameters mediating KBM transport. Considering Figure \ref{fig:sparc_density_prediction_transport}, the reason for this becomes apparent. First, it appears that $D_\mathrm{ped}$ is actually lower throughout the pedestal in this high density case than in the PRD. Pedestal values of $T_{e}$ and $\nabla T_{e}$ have both been decreased to represent typical higher density profiles. This decreases $D_\mathrm{neo}$, since that depends on $T_{e}$ strongly through the transport term, as mentioned above. $D_\mathrm{TG}$ increases slightly, since for the same input power, $\nabla T_{e}$ is lower, lowering $\chi_{e,\mathrm{eff}}$. These relatively low contributions to $D_\mathrm{ped}$ give very high $n_{e}^\mathrm{ped}$, especially as $n_{0}^\mathrm{sep}$ increases. 


As motivated in this paper, it is possible that at these high densities proposed, the additional RBM-driven transport channel may become important, and its absence in the prediction may in part help to explain the high $n_{e}^\mathrm{ped}$ values predicted above. Figure \ref{fig:sparc_density_prediction_transport} shows profiles of $D_\mathrm{RBM}$ computed from the final predicted $n_{e}$ (and input $T_{e}$) profile for the standard C-Mod plasma transport settings, using Equation \ref{eq:d_rbm_alphat}. While this transport channel is not included in the prediction and so is not self-consistently adjusted in the plasma transport, two things are clear from the figure. First, $D_\mathrm{RBM}$ is most active near the separatrix, where $\alpha_{t}$ is high. Second, $D_\mathrm{RBM}$ is considerably higher in the high density case, both than the other components of $D_\mathrm{ped}$ and than $D_\mathrm{RBM}$ in the PRD. Including a transport channel whose drive depends strongly on parameters related to decreased adiabaticity, which occurs at both high $n_{e}$ and low $T_{e}$, recovers the expectations that higher $n_{e}$ scenarios should experience larger particle transport, at least near the separatrix, than an equally-powered but lower density discharge.

Figure \ref{fig:sparc_rbm} shows the impact of including the RBM transport channel self-consistently on the prediction, for both the PRD (now in pink) and the higher density discharge (purple). This figure shows how the prediction changes from the standard C-Mod plasma transport settings, shown as thick, solid curves in Figure \ref{fig:sparc_density_prediction} when accounting for RBM transport using the model from Equation \ref{eq:d_rbm_alphat}. For the PRD, there is some difference in the $n_{e}^\mathrm{ped}$ prediction, as the RBM-driven transport weakens $\nabla n_{e}$ near the separatrix. Given that $\alpha_{t}$ is low for the PRD, this is likely an overestimate of the effect of the RBM. While the RBM expressions suggested do not have as sharp an onset condition as the KBM, for example, it could be that a threshold implementation, whereby the RBM transport turns on at a critical $\alpha_{t}$ (or $k_\mathrm{RBM}$), may be more appropriate. Or if indeed, the KBM triggers RBM-coupling at high $\alpha_{t}$, some more sophisticated reduced model including both KBM and RBM instability may be more appropriate. Regardless, for the high density discharge, where $\alpha_{t}$ is high, the difference in the prediction is quite drastic. Including this transport channel lowers $n_{e}^\mathrm{ped}$ considerably, down to $n_{e}^\mathrm{ped} = 5.1 \times 10^{20}$ m$^{-3}$, only 25\% above $n_{e}^\mathrm{sep}$, yielding a fairly flat pedestal profile. Whether this now overestimates the contribution from $D_\mathrm{RBM}$ is of course also uncertain. But, a quick glance back to Figure \ref{fig:ped_sep_neutrals_combined}, shows that once a high-density regime is accessed, $n_{e}^\mathrm{ped}$ may reach a naturally limited value, and it may not be much higher than $n_{e}^\mathrm{sep}$, especially at the highest values of $n_{e}^\mathrm{sep}$. Indeed, this prediction gives a ratio $\frac{n_{e}^\mathrm{sep}}{n_{e}^\mathrm{ped}} \approx 0.8$, on the upper end, but still within the range, of values for the C-Mod EDA pedestals analyzed in earlier sections.

\section{Conclusions}
\label{sec:conclusions}

Using a high quality archival dataset on Alcator C-Mod and state-of-the art modeling, this paper presents analysis suggesting the desirability of a high-density scenario like the EDA H-mode for reactor operation. The dataset is used as a basis for studying edge plasma and neutral transport and fluctuations properties and across ELMy and EDA H-mode regimes. It makes contact with and expands on pedestal models, using the findings to make predictions for the edge density of SPARC. On Alcator C-Mod, the standard operating scenario was the EDA H-mode, favored by the high density characteristic of this machine. ELMs were rare and could only be provoked through density control and modified shaping. On the particular run day examined here, once the density was reduced beyond a critical value, the H-mode transitioned away from the EDA and into an ELMy H-mode. The analysis presented here shows that as the density continued to drop, so did the size of the ELMs. This transition away from the EDA H-mode also coincides with a change in mode activity and transport properties influencing edge densities. In the EDA, the pedestal and its gradients do not respond to large changes in sources, presumably regulated instead by elevated particle transport. ELMy H-modes, on the other hand, experience rapid changes in densities and gradients in response to changing sources.


To build understanding of fluctuations and turbulent transport in these regimes, Fourier spectra from PCI measurements were evaluated. Attention was placed on assessing another trademark of the EDA H-mode -- the QCM. The evolution of its strength is tracked within the EDA H-mode, as well as from the mid-frequency elevated, although more broadband, fluctuations in ELMy H-modes. From this fluctuation analysis alone, it is not possible to isolate one parameter that is most important in the drive of the QCM, but it does speculate that its existence depends on high edge $\beta$ (high pressure and/or pressure gradient), as well as high edge $\nu^{*}$, the requisite ingredients to add EM character to any number of resistive modes. In the EDA H-mode, background fluctuation levels closely track these same dimensionless parameters drive changes, but in ELMy H-modes, they are nearly constant. The trends presented here may serve to provide an experimental comparison on which to build understanding of how fluctuation-driven particle transport may interact with MHD modes to produce small or large ELMs or ultimately avoid them.

Motivated by these experimental observations, the latter half of this work examined leading models for pedestal prediction on these diverse C-Mod H-modes, using them for comparison and suggesting expansions where possible. It validated the Saarelma-Connor model for the first time on Alcator C-Mod pedestals, extending the validity range of the model to pedestal densities up to 3.0 $\times 10^{20}$ m$^{-3}$, above the expected operation range of next-step devices and pilot plants. Using reasonable estimates for a boundary condition on the neutral solution estimated by KN1D, the prediction for ELMy H-modes, up to $n_{e}^\mathrm{ped} = 2.0 \times 10^{20}$ m$^{-3}$, closely matched the experimental values. For EDA H-modes, stronger particle transport was necessary. This work has suggested two different forms for the additional transport channel driven by the RBM, motivated by recent database work with high quality particle transport measurements. For appropriately-chosen coefficients, this addition to the model improves the prediction across the experimental range of densities. EPED simulations were then carried out to evaluate the applicability of the KBM-PBM constraint on the pedestal structure across this range of H-modes. The model was found somewhat sensitive to changes in the ratio of separatrix to pedestal density, especially for peeling-limited pedestals and importantly, in the transition between the peeling and ballooning branches. When compared to experimental data, model agreement is found for large ELMy discharges, confirming a previous result. The model overpredicts the pressure for all small ELMy H-modes and many EDA H-modes, but notably, it underpredicts it for a non-negligible number of EDAs. This exercise has suggested that more work remains to understand how both turbulence and MHD stability interact to set edge profiles.

This work ends with predictions for the SPARC tokamak, whose goal is to produce scientific breakeven. To do so, it must integrate a high performance core with an edge solution that will promote divertor survivability. One of the keys to executing this goal lies in tailoring the pedestal profile. To this end, preliminary predictions are presented for pedestal density profiles in both the standard PRD scenario as well as a recently proposed high-density scenario without Type-I ELMs, using the Saarelma-Connor model and experimental insight into transport at high density. Predictions for the PRD are in line with those from previous EPED modeling of this scenario. The higher density H-mode is somewhat more sensitive to modeling assumptions. In particular, RBM-driven transport is found to strongly affect density gradients near the separatrix, limiting the achievable pedestal density. Given its already high separatrix density, an only somewhat higher pedestal density may actually be beneficial from the perspective of finding a core solution without significant radiative losses. Of course, work must be done to find such a solution, as well as to improve confidence in these models and reduce uncertainties in their inputs.

Due to the high potential for ELM-related damage to plasma facing components in burning plasmas, it is imperative to understand access to and improve predictive capabilities of performance in regimes free of Type-I ELMs. Throughout its operation, Alcator C-Mod has proven highly useful for studying access to and performance of such regimes. This work represents one small part of a large body of active work to understand how to avoid large ELMs, while maximizing performance. As the experimental analysis and modeling of this paper suggests, the EDA H-mode, and possibly other high density regimes free of Type-I ELMs sharing characteristics with the EDA, may not only be necessary but also attractive operational regimes for a fusion power plant. 

\section*{Acknowledgments}
The authors would like to thank R.M. McDermott for helpful conversations on physical interpretation of the experimental data and G. Birkenmeier for useful discussions on and suggestions for analysis of fluctuation measurements. This work was supported in part by US DOE Awards DE-SC0021629, DE-SC0014264, DE-SC0023289 and DE-SC0007880 and by Commonwealth Fusion Systems. It has also has been carried out within the framework of the EUROfusion Consortium, funded by the European Union via the Euratom Research and Training Programme (Grant Agreement No 101052200 — EUROfusion). Views and opinions expressed are however those of the author(s) only and do not necessarily reflect those of the European Union or the European Commission. Neither the European Union nor the European Commission can be held responsible for them.

\appendix
\section{KN1D simulations for $n_{0}^\mathrm{sep}$ parameterization}
\label{sec:kn1d_sims}

\begin{figure}
\centering
\includegraphics[width=0.9\columnwidth]{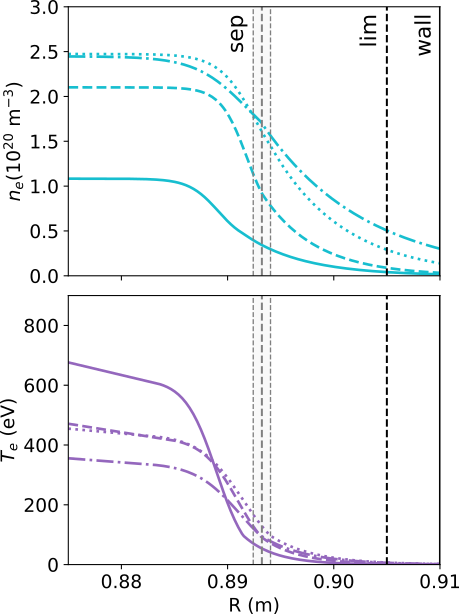}
\caption{Characteristic $n_{e}$ (top) and $T_{e}$ (bottom) profiles for ELMy-EDA dataset from ``moving-average'' technique for 1st, 4th, 7th, and 10th pressure range shown in Figure \ref{fig:dataset_moving_averages} as dash-dotted, dotted, dashed, and solid curves. Mean $p_{0}^\mathrm{wall}$ is 0.037, 0.087, 0.23, and 0.52 mTorr for the pressure range corresponding to these characteristic profiles, respectively. Radial positions of ``wall,'' limiter, and separatrix ranges are shown in solid black, dashed black, and gray lines and shaded regions.}
\label{fig:characteristic_profiles_for_kn1d}
\end{figure}


The KN1D simulations presented in this Appendix are intended to provide a neutral solution between the wall, where the measurement of $p_{0}^\mathrm{OMP}$ is readily available, and the separatrix, where $n_{0}^\mathrm{sep}$ is a required input to the Saarelma-Connor model. KN1D is 1D in space and 2D in velocity. It models neutral transport of atomic and molecular hydrogen, including a number of plasma-neutral and neutral-neutral collision mechanisms known to be important for neutral transport in the edge. KN1D calculates neutral dynamics on a fixed plasma background, and it does not evolve the plasma solution in response to changes in the neutral source. The code has been used to study edge neutrals extensively on Alcator C-Mod \cite{hughes_advances_2006, hughes_edge_2007}, as well as on JET \cite{maddison_dimensionless_2009}. A detailed comparison between KN1D and the neutrals model of Saarelma and Connor across the pedestal is now underway \cite{dunsmore_validating_2025}.

Since it is a 1D code, KN1D requires few inputs. It uses a user-supplied plasma radial grid, $r$, on which the plasma background is provided, defined by $n_{e}(r)$, $T_{e}(r)$, and $T_{i}(r)$. Furthermore, it takes in the position of the separatrix and optionally, a limiter, defined by $r_\mathrm{sep}$ and $r_\mathrm{lim}$, respectively, which model the scenario where the magnetic divertor or primary limiter is remote from a local limiter, as was the case for these C-Mod plasmas. One can also specify a connection length, which is applied in the limiter shadow to ensure that plasma flow to the limiter surfaces is proportional to the local sound speed, scaled by the connection length. Furthermore, the code assumes a fully recycling boundary condition, achieved by scaling the integral of the plasma flow to the limiter sides to give a net zero mass flux to and from the combined limiters and wall. Finally, and most importantly for this exercise, the boundary condition for the molecular hydrogen distribution function is given by a specified wall pressure, $p_{0}^\mathrm{wall}$. Functionally, this serves to scale the magnitude of the distribution function of neutrals launched towards the plasma. For the purposes of this simulation exercise, however, $p_{0}^\mathrm{wall}$ is available experimentally. It is set to $p_{0}^\mathrm{OMP}$ and, when combined with a known plasma background, yields the requisite $n_{0}^\mathrm{sep}$ to constrain the Saarelma-Connor model at the separatrix.

\begin{figure}
\centering
\includegraphics[width=\columnwidth]{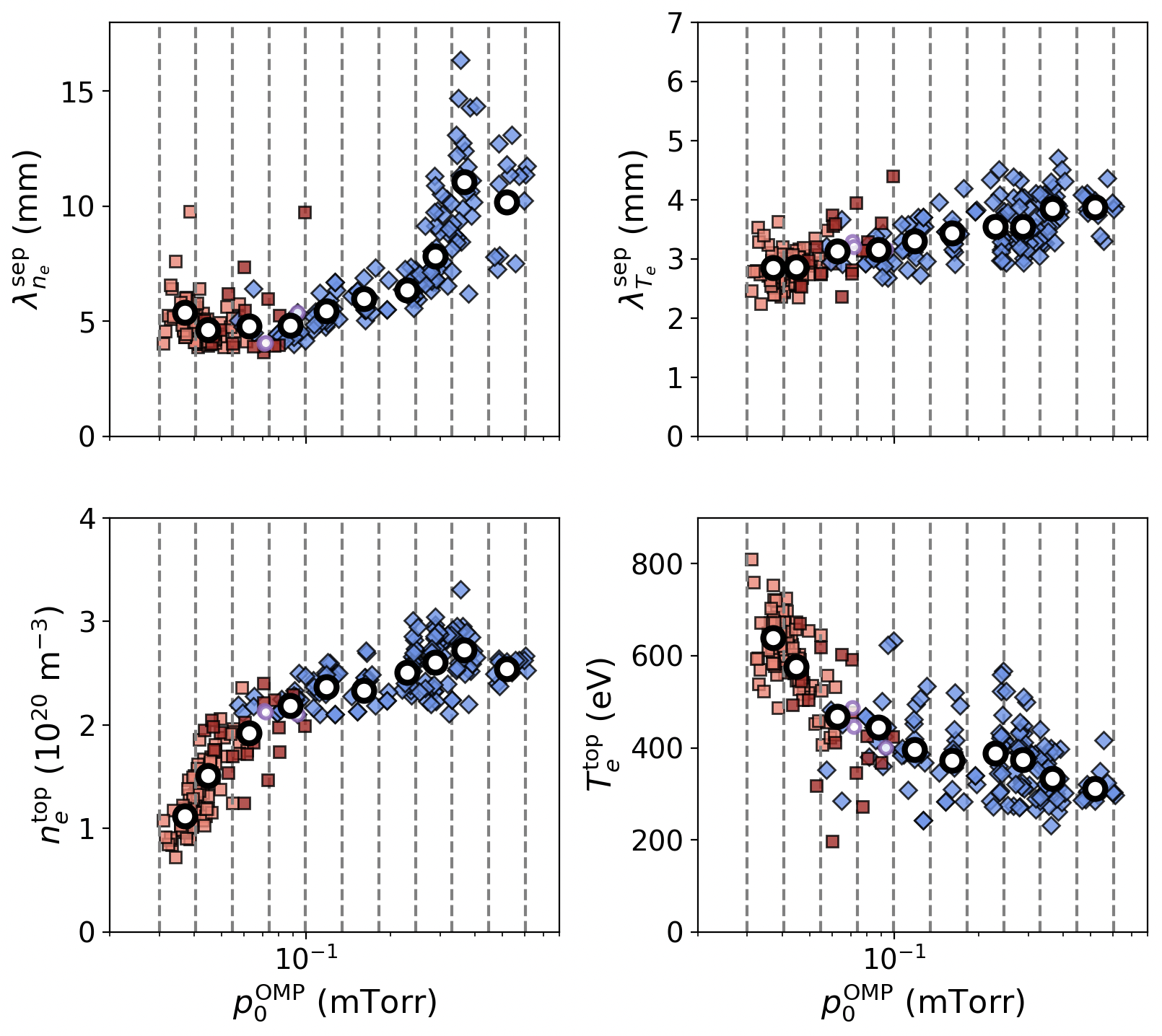}
\caption{Separatrix gradient scale lengths (top) and pedestal top values (bottom) for $n_{e}$ (left) and $T_{e}$ (right) plotted against $p_{0}^\mathrm{OMP}$, showing experimental values for ELMy-EDA dataset in previous color scheme and overlaying ``moving averages'' in open white circles. Vertical dashed gray lines show ranges in $p_{0}^\mathrm{OMP}$ over which moving averages are computed.}
\label{fig:dataset_moving_averages}
\end{figure}

To get a sense for the functional mapping between $p_{0}^\mathrm{OMP}$ and $n_{0}^\mathrm{sep}$ in this dataset, a modest number of KN1D runs are prepared using a range of plasma parameters from the larger data set. Plasma profiles used as KN1D inputs consist of characteristic profiles constructed by joining an mtanh fit inside the separatrix with an exponential decay fit outside. To build these profiles, information about the pedestal top and bottom, its width, the position of the pedestal and separatrix independently, as well as the SOL decay lengths are required. These profile parameters are taken directly from the experimental database using a moving-average procedure, as explained below. Pedestal structure parameters come from the mtanh fits, while information about the separatrix position and SOL decay lengths is taken from the exponential decay fits. The full set of average variables required to construct characteristic profiles is similar for both $n_{e}$ and $T_{e}$, except inside the pedestal top. For the $n_{e}$ profile, for example, values of $n_{e}^\mathrm{top}$, $n_{e}^\mathrm{bot}$, the center of the pedestal, and its width are sufficient to construct a characteristic mtanh pedestal profile. This tanh profile is then aligned to the separatrix by shifting it to yield $n_{e}^\mathrm{sep}$ at $R_\mathrm{sep}$. Note, however, that since the mtanh profile was constructed prior to the shift, information about the position of the center of the mtanh relative to the separatrix is preserved. 

At the separatrix, the mtanh profile is truncated, and an exponential decay given by $\lambda_{n_{e}}$ is applied. This exponential fit is designed to more appropriately characterize the SOL plasma than the mtanh, which can artificially fill the SOL with plasma, strongly impacting the neutral penetration characteristics. Depending on the interplay between the specified pedestal width, the distance between the center to the separatrix, and the specified SOL decay length, this profile construction technique can, and often does, introduce discontinuities in the gradient at the separatrix\footnote{A more clever fit function to accurately represent the pedestal and SOL plasma while ensuring a continuous derivative in the profile is likely possible, but KN1D only considers the plasma profiles, not their gradients. Since the profiles are continuous, this profile-building approach is deemed adequate for this exercise.}. The same approach is taken for $T_{e}$ profiles with the added detail that a linear function is specified inside the pedestal top, using similarly chosen mean $\nabla T_{e}^\mathrm{ped}$ values. While this more faithfully reproduces the shape of $T_{e}$ pedestal profiles, it is found to have no significant impact on the mapping between $p_{0}^\mathrm{OMP}$ and $n_{0}^\mathrm{sep}$. Four example profiles generated with this technique are shown in Figure \ref{fig:characteristic_profiles_for_kn1d}, with $n_{e}$ at top and $T_{e}$ at bottom. Included in this figure are also the position of the wall from which neutrals are launched in KN1D as explained below, the limiter (lim) position, and the mean position of the separatrix (sep), including an error band, since the position of the separatrix is not fixed with reference to the vessel. An additional fall-off outside of the limiter location, $R_\mathrm{lim}$, might be more appropriate to characterize the plasma density near the wall, such that the results presented here might overestimate the particle content behind the limiter.

The set of ten input profiles for KN1D are designed to span the full range of $p_{0}^\mathrm{OMP}$ in the OS. This is done by choosing a set of 10 geometrically-spaced intervals in $p_{0}^\mathrm{OMP}$, ranging between $p_{0}^\mathrm{OMP} = 0.03 - 0.6$ mTorr. For each of these intervals, average values of the above pedestal and SOL structure parameters are used to construct the characteristic profiles. For each of these structure variables, the procedure yields a ``moving average'' as a function of $p_{0}^\mathrm{OMP}$, at discretely chosen intervals of $p_{0}^\mathrm{OMP}$. This approach is thought more robust to natural scatter in the data than just picking individual fits arbitrarily. Examples of this procedure are shown in Figure \ref{fig:dataset_moving_averages}, showing the moving average for separatrix plasma decay lengths from the exponential decay fit at top and pedestal top values from the mtanh fit at bottom, for $n_{e}$ at left and $T_{e}$ at right. For these variables, $p_{0}^\mathrm{OMP}$ conveniently appears an appropriate organizing variable on which to compute these averages, except perhaps at high $n_{e}$.

\begin{figure}
\centering
\includegraphics[width=0.9\columnwidth]{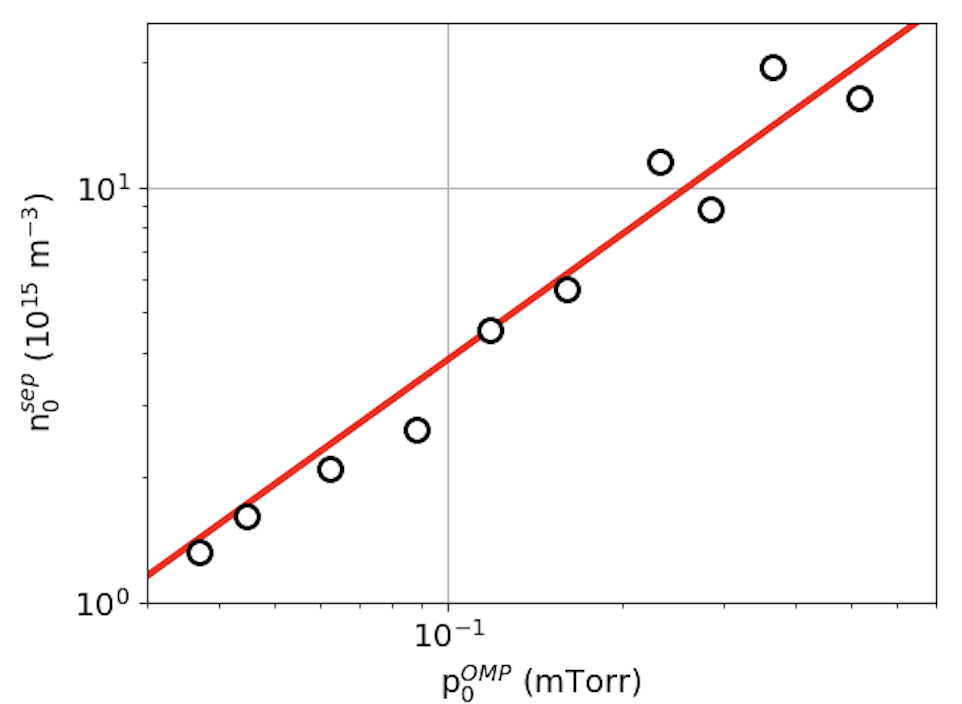}
\caption{$n_{0}^\mathrm{sep}$ resulting from KN1D using scan in $p_{0}^\mathrm{OMP}$ and corresponding characteristic $n_{e}$ and $T_{e}$ profiles plotted against $p_{0}^\mathrm{OMP}$ used in the simulation, shown as white circles. Best linear fit, $n_{0}^\mathrm{sep}\mathrm{[10^{15} m^{-3}]} = 38.5 p_{0}^\mathrm{OMP} \mathrm{[mTorr]}$ is shown in red.}
\label{fig:n0_pwall_ratio}
\end{figure}

These ten characteristic $n_{e}$ and $T_{e}$ profiles (including $T_{i}$ under the assumption that $T_{e} = T_{i}$, an assumption that can and should likely be relaxed at low $n_{e}$), as well as the mean $p_{0}^\mathrm{OMP}$ for each interval, are used to run KN1D. The code calculates and returns self-consistent velocity space distributions of atomic and molecular hydrogen across the radial grid, whose moments can be computed and combined with atomic rates to estimate neutral densities, temperatures, and fluxes. For this exercise, it is the atomic neutral density profile, $n_{0}(r)$, that is of interest. In particular, the values of $n_{0}^\mathrm{sep}=n_{0}(r=r_\mathrm{sep})$ are extracted for each set of inputs. The results are plotted against the input $p_{0}^\mathrm{OMP}$, as shown in Figure \ref{fig:n0_pwall_ratio}. The data are then fit to establish the desired functional mapping between $p_{0}^\mathrm{OMP}$ and $n_{0}^\mathrm{sep}$. The relation is rather linear, except possibly at the highest $p_{0}$. One could imagine that at high $p_{0}^\mathrm{OMP}$, when $n_{e}^\mathrm{sep}$ and $\lambda_{n_{e}}^\mathrm{sep}$ are both large, $n_{e}^\mathrm{SOL}$ may act to screen neutrals, slightly saturating the value of $n_{0}^\mathrm{sep}$. A finer scan in $p_{0}^\mathrm{OMP}$ at high $p_{0}^\mathrm{OMP}$ would help to resolve this, but for the purposes of testing an already reduced model, the qualitative observation that $n_{0}^\mathrm{sep} \sim p_{0}^\mathrm{OMP}$ in this dataset is sufficient to specify boundary conditions for the Saarelma-Connor model. In general, the KN1D results suggest that $n_{0}^\mathrm{sep}$ can vary considerably, not altogether surprising given the large variation in $p_{0}^\mathrm{OMP}$ (and $n_{e}$) in this dataset. For $p_{0}^\mathrm{OMP} > 0.1$ mTorr, the pressure range at which EDAs are found, $n_{0}^\mathrm{sep}$ can range from $4 \times 10^{15}$ m$^{-3}$ up to $3 \times 10^{16}$ m$^{-3}$. This lower bound for EDAs is slightly lower than previously-reported values for EDAs but is consistent with the different shape and accompanying more closed divertor. For ELMy discharges with $p_{0}^\mathrm{OMP} < 0.1$ mTorr, KN1D calculates $n_{0}^\mathrm{sep}$ as low as $10^{15}$ m$^{-3}$, which matches the scalar value of $n_{0}^\mathrm{sep}$ used in Figure \ref{fig:results_elmy_eda_transport} and in previous validation efforts of the Saarelma-Connor model.


\section{Setup of EPED interpretive scans}
\label{sec:eped_details}

To provide a point of reference for ideal MHD pedestal stability predictions, the EPED code \cite{snyder_development_2009, snyder_eped_2012}, version 1.0, is run for a range of $n_{e}^\mathrm{ped}$ spanning the database in this work, and the predictions are then compared against the experimental data. EPED uses two separate constraints to make a prediction for the height and width of the pressure pedestal of Type-I ELMy H-modes. The first of these is a constraint imposed by KBMs. These can drive large amounts of transport in the pedestal and become active at a critical gradient in the poloidal beta, $\beta_{p} = \frac{p}{B_{p}^2/2\mu_{0}}$. Instead of the full gyrokinetic calculation for the KBM, this constraint is supplanted in EPED1.0 with a pedestal width-height scaling. This scaling is given by $\Delta_{p} = C\sqrt{\beta_{p}^\mathrm{ped}}$, with $\Delta_{p}$ the pressure pedestal width in $\psi_{n}$ and $\beta_{p}^\mathrm{ped}$ the pedestal height in terms of the normalized $\beta_{p}$ evaluated at the pedestal top. The proportionality coefficient, $C$, has been determined empirically and is typically $O(10^{-1})$. This KBM scaling has successfully described the pedestal structure on a number of devices, the majority of which are also remarkably described by a common value of $C = 0.076$ \cite{urano_dimensionless_2008, snyder_pedestal_2009, kirk_comparison_2009, maggi_pedestal_2010, beurskens_h-mode_2011, groebner_improved_2013}, with a slight variation in $C$ for C-Mod pedestals \cite{walk_characterization_2012, hughes_pedestal_2013, snyder_pedestal_2009} and a different width-height scaling on NSTX \cite{diallo_progress_2013}, which has since been reconciled \cite{parisi_kinetic-ballooning-bifurcation_2024}. As was shown in Figure \ref{fig:kbm_broadening}, for this particular dataset of C-Mod discharges, the standard value in EPED1.0 of $C = 0.076$ describes the structure of many discharges well, and is thus used in the simulations. This scaling is used to construct a family of KBM-limited profiles, which are then combined with a constraint on PBMs. These modes are global MHD modes that act across the extent of the pedestal, limiting both its height and width. Linear stability to PBMs is quite computationally tractable and is done in EPED with the code ELITE \cite{wilson_numerical_2002}. While all PBMs have some peeling and ballooning character, in general, modes with lower toroidal mode number, $n$, are more peeling-dominated and are driven unstable at high current, while higher-$n$ modes are more ballooning-dominated and are driven unstable at high pressure gradient. 

EPED requires nine input parameters: $I_{P}$, $B_{t}$, $R_{0}$, $a$, $\kappa$, $\delta$, $\beta_{N}$, $Z_\mathrm{eff}^\mathrm{ped}$, and $n_{e}^\mathrm{ped}$, with $\beta_{N}$ the global normalized beta, $Z_\mathrm{eff}^\mathrm{ped}$ the effective charge at the pedestal top, and all other quantities previously defined in the body of this work. Note that the requirement of $n_{e}^\mathrm{ped}$ as an input is one of the limitations of EPED-like pedestal predictions and exactly the reason a good deal of effort has been devoted to independently predict $n_{e}^\mathrm{ped}$, as detailed in Section \ref{sec:density_model_validation}. In this database, there is little variation in most inputs, including $\beta_{N}$, except for $n_{e}^\mathrm{ped}$. The modeling strategy adopted is thus to use only one set of scalar inputs, determined by taking the mean value of the experimental parameters to be used as inputs across the database, and varying only $n_{e}^\mathrm{ped}$. In this way, $n_{e}^\mathrm{ped}$ is scanned from 0.6 -- 2.6 $\times 10^{20}$ m$^{-3}$, covering the full range of plasma densities realized experimentally. This approach  mirrors the approach typically used when running EPED in a predictive mode, instead of interpretively, as is done here. Scanning $n_{e}^\mathrm{ped}$ in this way provides a reference, if not a direct experimental validation, for how the constraints in EPED would suggest $p^\mathrm{ped}$ would scale with $n_{e}^\mathrm{ped}$ experimentally. One can then read off whether the experimental realization matches the expectation from the theory or whether additional physical constraints need be invoked to explain the experiment. Specifically, from a scan in $n_{e}^\mathrm{ped}$, one can determine whether a pedestal will be primarily peeling-limited, in which case $p^\mathrm{ped}$ (and $\Delta_{p}$) would correlate positively with $n_{e}^\mathrm{ped}$, or ballooning-limited, in which it would decrease with increasing $n_{e}^\mathrm{ped}$. 



While all primary inputs into EPED are relatively fixed, one last parameter that is typically left fixed but can be modified is the ratio of separatrix to pedestal density, $\frac{n_{e}^\mathrm{sep}}{n_{e}^\mathrm{ped}}$. This value is typically set to 0.25, and for many H-modes, especially Type-I ELMy H-modes, this is a fairly good estimate. For the dataset presented here, however, $n_{e}^\mathrm{sep}$ and $n_{e}^\mathrm{ped}$ can vary significantly, and also independently, such their ratio also varies significantly. Indeed, it can range anywhere from just below 0.4 to just above 0.8. To understand the sensitivity of the code's predictions to this ratio, three separate density scans are performed at three different values of $\frac{n_{e}^\mathrm{sep}}{n_{e}^\mathrm{ped}}$: 0.4, 0.6, and 0.8.

\section*{References}

\bibliographystyle{unsrt}
\bibliography{references}

\end{document}